\documentclass[twocolumn]{aastex61}
\usepackage[utf8]{inputenc}
\usepackage{colortbl}
\usepackage{verbatim}
\usepackage[maxfloats=25,morefloats=7]{morefloats}

\received{XXXXXX XX, XXXX}
\revised{XXXXXX XX, XXXX}
\accepted{XXXXXX XX, XXXX}
\submitjournal{AJ}

\shorttitle{TNOs and Centaurs from DES: astrometry and occultations}
\shortauthors{Banda-Huarca, M.V. et al.}

\begin{document}

\newcommand{\notification}[2]{\textcolor{red}{#1} \textcolor{green}{#2}}

\vspace*{-\headsep}\vspace*{\headheight}
{\footnotesize \hfill FERMILAB-PUB-18-494}\\
\vspace*{-\headsep}\vspace*{\headheight}
{\footnotesize \hfill DES-2017-0300}

\title{Astrometry and occultation predictions to transneptunian and Centaur objects observed within the Dark Energy Survey }

\correspondingauthor{Banda-Huarca, M.V.}
\email{martin.banda@linea.gov.br}

\author{M.~V.~Banda-Huarca}
\affiliation{Observat\'orio Nacional, Rua Gal. Jos\'e Cristino 77, Rio de Janeiro, RJ - 20921-400, Brazil}
\affiliation{Laborat\'orio Interinstitucional de e-Astronomia - LIneA, Rua Gal. Jos\'e Cristino 77, Rio de Janeiro, RJ - 20921-400, Brazil}

\author[0000-0002-1642-4065]{J.~I.~B.~Camargo}
\affiliation{Observat\'orio Nacional, Rua Gal. Jos\'e Cristino 77, Rio de Janeiro, RJ - 20921-400, Brazil}
\affiliation{Laborat\'orio Interinstitucional de e-Astronomia - LIneA, Rua Gal. Jos\'e Cristino 77, Rio de Janeiro, RJ - 20921-400, Brazil}

\author[0000-0002-2193-8204]{J.~Desmars}
\affiliation{LESIA, Observatoire de Paris, PSL Research University, CNRS, Sorbonne Universitées, UPMC Univ.  Paris 06, Univ.  Paris Diderot, Sorbonne Paris Cité, France}

\author[0000-0003-2120-1154]{R.~L.~C.~Ogando}
\affiliation{Observat\'orio Nacional, Rua Gal. Jos\'e Cristino 77, Rio de Janeiro, RJ - 20921-400, Brazil}
\affiliation{Laborat\'orio Interinstitucional de e-Astronomia - LIneA, Rua Gal. Jos\'e Cristino 77, Rio de Janeiro, RJ - 20921-400, Brazil}

\author[0000-0003-1690-5704]{R.~Vieira-Martins}
\affiliation{Observat\'orio Nacional, Rua Gal. Jos\'e Cristino 77, Rio de Janeiro, RJ - 20921-400, Brazil}
\affiliation{Laborat\'orio Interinstitucional de e-Astronomia - LIneA, Rua Gal. Jos\'e Cristino 77, Rio de Janeiro, RJ - 20921-400, Brazil}

\author[0000-0002-8211-0777]{M.~Assafin}
\affiliation{Observat\'orio do Valongo, Ladeira do Pedro Antonio, 43 - Saude 20.080-090, Rio de Janeiro, RJ - Brazil}
\affiliation{Laborat\'orio Interinstitucional de e-Astronomia - LIneA, Rua Gal. Jos\'e Cristino 77, Rio de Janeiro, RJ - 20921-400, Brazil}

\author{L.~N.~da Costa}
\affiliation{Observat\'orio Nacional, Rua Gal. Jos\'e Cristino 77, Rio de Janeiro, RJ - 20921-400, Brazil}
\affiliation{Laborat\'orio Interinstitucional de e-Astronomia - LIneA, Rua Gal. Jos\'e Cristino 77, Rio de Janeiro, RJ - 20921-400, Brazil}

\author[0000-0002-7555-2956]{G.~M.~Bernstein}
\affiliation{Department of Physics and Astronomy, University of Pennsylvania, Philadelphia, PA 19104, USA}

\author[0000-0002-4802-3194]{M.~Carrasco~Kind}
\affiliation{Department of Astronomy, University of Illinois at Urbana-Champaign, 1002 W. Green Street, Urbana, IL 61801, USA}
\affiliation{National Center for Supercomputing Applications, 1205 West Clark St., Urbana, IL 61801, USA}

\author[0000-0001-8251-933X]{A.~Drlica-Wagner}
\affiliation{Fermi National Accelerator Laboratory, P. O. Box 500, Batavia, IL 60510, USA}
\affiliation{Kavli Institute for Cosmological Physics, University of Chicago, Chicago, IL 60637, USA}

\author[0000-0001-5712-3042]{R.~Gomes}
\affiliation{Observat\'orio Nacional, Rua Gal. Jos\'e Cristino 77, Rio de Janeiro, RJ - 20921-400, Brazil}
\affiliation{Laborat\'orio Interinstitucional de e-Astronomia - LIneA, Rua Gal. Jos\'e Cristino 77, Rio de Janeiro, RJ - 20921-400, Brazil}

\author{M. M. Gysi}
\affiliation{Federal University of Technology - Paraná (UTFPR/DAFIS), Av. Sete de Setembro 3165, 80230-901, Curitiba, Brazil}
\affiliation{Laborat\'orio Interinstitucional de e-Astronomia - LIneA, Rua Gal. Jos\'e Cristino 77, Rio de Janeiro, RJ - 20921-400, Brazil}

\author[0000-0003-2311-2438]{F.~Braga-Ribas}
\affiliation{Federal University of Technology - Paraná (UTFPR/DAFIS), Av. Sete de Setembro 3165, 80230-901, Curitiba, Brazil}
\affiliation{Laborat\'orio Interinstitucional de e-Astronomia - LIneA, Rua Gal. Jos\'e Cristino 77, Rio de Janeiro, RJ - 20921-400, Brazil}
\affiliation{Observat\'orio Nacional, Rua Gal. Jos\'e Cristino 77, Rio de Janeiro, RJ - 20921-400, Brazil}

\author[0000-0001-9856-9307]{M.~A.~G.~Maia}
\affiliation{Observat\'orio Nacional, Rua Gal. Jos\'e Cristino 77, Rio de Janeiro, RJ - 20921-400, Brazil}
\affiliation{Laborat\'orio Interinstitucional de e-Astronomia - LIneA, Rua Gal. Jos\'e Cristino 77, Rio de Janeiro, RJ - 20921-400, Brazil}

\author[0000-0001-6942-2736 ]{D.~W.~Gerdes}
\affiliation{Department of Physics, University of Michigan, Ann Arbor, MI 48109, USA}
\affiliation{Department of Astronomy, University of Michigan, Ann Arbor, MI 48109, USA}

\author{S.~Hamilton}
\affiliation{Department of Physics, University of Michigan, Ann Arbor, MI 48109, USA}

\author{W.~Wester}
\affiliation{Fermi National Accelerator Laboratory, P. O. Box 500, Batavia, IL 60510, USA}

\author{T.~M.~C.~Abbott}
\affiliation{Cerro Tololo Inter-American Observatory, National Optical Astronomy Observatory, Casilla 603, La Serena, Chile}

\author[0000-0003-2063-4345]{F.~B.~Abdalla}
\affiliation{Department of Physics \& Astronomy, University College London, Gower Street, London, WC1E 6BT, UK}
\affiliation{Department of Physics and Electronics, Rhodes University, PO Box 94, Grahamstown, 6140, South Africa}

\author[0000-0002-7069-7857]{S.~Allam}
\affiliation{Fermi National Accelerator Laboratory, P. O. Box 500, Batavia, IL 60510, USA}

\author{S.~Avila}
\affiliation{Institute of Cosmology \& Gravitation, University of Portsmouth, Portsmouth, PO1 3FX, UK}

\author{E.~Bertin}
\affiliation{CNRS, UMR 7095, Institut d'Astrophysique de Paris, F-75014, Paris, France}
\affiliation{Sorbonne Universit\'es, UPMC Univ Paris 06, UMR 7095, Institut d'Astrophysique de Paris, F-75014, Paris, France}

\author{D.~Brooks}
\affiliation{Department of Physics \& Astronomy, University College London, Gower Street, London, WC1E 6BT, UK}

\author{E.~Buckley-Geer}
\affiliation{Fermi National Accelerator Laboratory, P. O. Box 500, Batavia, IL 60510, USA}

\author{D.~L.~Burke}
\affiliation{Kavli Institute for Particle Astrophysics \& Cosmology, P. O. Box 2450, Stanford University, Stanford, CA 94305, USA}
\affiliation{SLAC National Accelerator Laboratory, Menlo Park, CA 94025, USA}

\author[0000-0003-3044-5150]{A.~Carnero~Rosell}
\affiliation{Laborat\'orio Interinstitucional de e-Astronomia - LIneA, Rua Gal. Jos\'e Cristino 77, Rio de Janeiro, RJ - 20921-400, Brazil}
\affiliation{Observat\'orio Nacional, Rua Gal. Jos\'e Cristino 77, Rio de Janeiro, RJ - 20921-400, Brazil}

\author[0000-0002-3130-0204]{J.~Carretero}
\affiliation{Institut de F\'{\i}sica d'Altes Energies (IFAE), The Barcelona Institute of Science and Technology, Campus UAB, 08193 Bellaterra (Barcelona) Spain}

\author{C.~E.~Cunha}
\affiliation{Kavli Institute for Particle Astrophysics \& Cosmology, P. O. Box 2450, Stanford University, Stanford, CA 94305, USA}

\author{C.~Davis}
\affiliation{Kavli Institute for Particle Astrophysics \& Cosmology, P. O. Box 2450, Stanford University, Stanford, CA 94305, USA}

\author[0000-0001-8318-6813]{J.~De~Vicente}
\affiliation{Centro de Investigaciones Energ\'eticas, Medioambientales y Tecnol\'ogicas (CIEMAT), Madrid, Spain}

\author[0000-0002-8357-7467]{H.~T.~Diehl}
\affiliation{Fermi National Accelerator Laboratory, P. O. Box 500, Batavia, IL 60510, USA}

\author{P.~Doel}
\affiliation{Department of Physics \& Astronomy, University College London, Gower Street, London, WC1E 6BT, UK}

\author{P.~Fosalba}
\affiliation{Institut d'Estudis Espacials de Catalunya (IEEC), 08193 Barcelona, Spain}
\affiliation{Institute of Space Sciences (ICE, CSIC),  Campus UAB, Carrer de Can Magrans, s/n,  08193 Barcelona, Spain}

\author[0000-0003-4079-3263]{J.~Frieman}
\affiliation{Fermi National Accelerator Laboratory, P. O. Box 500, Batavia, IL 60510, USA}
\affiliation{Kavli Institute for Cosmological Physics, University of Chicago, Chicago, IL 60637, USA}

\author[0000-0002-9370-8360]{J.~Garc\'ia-Bellido}
\affiliation{Instituto de Fisica Teorica UAM/CSIC, Universidad Autonoma de Madrid, 28049 Madrid, Spain}

\author{E.~Gaztanaga}
\affiliation{Institut d'Estudis Espacials de Catalunya (IEEC), 08193 Barcelona, Spain}
\affiliation{Institute of Space Sciences (ICE, CSIC),  Campus UAB, Carrer de Can Magrans, s/n,  08193 Barcelona, Spain}

\author{D.~Gruen}
\affiliation{Kavli Institute for Particle Astrophysics \& Cosmology, P. O. Box 2450, Stanford University, Stanford, CA 94305, USA}
\affiliation{SLAC National Accelerator Laboratory, Menlo Park, CA 94025, USA}

\author{R.~A.~Gruendl}
\affiliation{Department of Astronomy, University of Illinois at Urbana-Champaign, 1002 W. Green Street, Urbana, IL 61801, USA}
\affiliation{National Center for Supercomputing Applications, 1205 West Clark St., Urbana, IL 61801, USA}

\author[0000-0003-3023-8362]{J.~Gschwend}
\affiliation{Observat\'orio Nacional, Rua Gal. Jos\'e Cristino 77, Rio de Janeiro, RJ - 20921-400, Brazil}
\affiliation{Laborat\'orio Interinstitucional de e-Astronomia - LIneA, Rua Gal. Jos\'e Cristino 77, Rio de Janeiro, RJ - 20921-400, Brazil}

\author[0000-0003-0825-0517]{G.~Gutierrez}
\affiliation{Fermi National Accelerator Laboratory, P. O. Box 500, Batavia, IL 60510, USA}

\author{W.~G.~Hartley}
\affiliation{Department of Physics \& Astronomy, University College London, Gower Street, London, WC1E 6BT, UK}
\affiliation{Department of Physics, ETH Zurich, Wolfgang-Pauli-Strasse 16, CH-8093 Zurich, Switzerland}

\author{D.~L.~Hollowood}
\affiliation{Santa Cruz Institute for Particle Physics, Santa Cruz, CA 95064, USA}

\author{K.~Honscheid}
\affiliation{Center for Cosmology and Astro-Particle Physics, The Ohio State University, Columbus, OH 43210, USA}
\affiliation{Department of Physics, The Ohio State University, Columbus, OH 43210, USA}

\author{D.~J.~James}
\affiliation{Harvard-Smithsonian Center for Astrophysics, Cambridge, MA 02138, USA}

\author[0000-0003-0120-0808]{K.~Kuehn}
\affiliation{Australian Astronomical Observatory, North Ryde, NSW 2113, Australia}

\author[0000-0003-2511-0946]{N.~Kuropatkin}
\affiliation{Fermi National Accelerator Laboratory, P. O. Box 500, Batavia, IL 60510, USA}

\author[0000-0002-1372-2534]{F.~Menanteau}
\affiliation{Department of Astronomy, University of Illinois at Urbana-Champaign, 1002 W. Green Street, Urbana, IL 61801, USA}
\affiliation{National Center for Supercomputing Applications, 1205 West Clark St., Urbana, IL 61801, USA}

\author{C.~J.~Miller}
\affiliation{Department of Astronomy, University of Michigan, Ann Arbor, MI 48109, USA}
\affiliation{Department of Physics, University of Michigan, Ann Arbor, MI 48109, USA}

\author[0000-0002-6610-4836]{R.~Miquel}
\affiliation{Instituci\'o Catalana de Recerca i Estudis Avan\c{c}ats, E-08010 Barcelona, Spain}
\affiliation{Institut de F\'{\i}sica d'Altes Energies (IFAE), The Barcelona Institute of Science and Technology, Campus UAB, 08193 Bellaterra (Barcelona) Spain}

\author[0000-0002-2598-0514]{A.~A.~Plazas}
\affiliation{Jet Propulsion Laboratory, California Institute of Technology, 4800 Oak Grove Dr., Pasadena, CA 91109, USA}

\author[0000-0002-9328-879X]{A.~K.~Romer}
\affiliation{Department of Physics and Astronomy, Pevensey Building, University of Sussex, Brighton, BN1 9QH, UK}

\author[0000-0002-9646-8198]{E.~Sanchez}
\affiliation{Centro de Investigaciones Energ\'eticas, Medioambientales y Tecnol\'ogicas (CIEMAT), Madrid, Spain}

\author{V.~Scarpine}
\affiliation{Fermi National Accelerator Laboratory, P. O. Box 500, Batavia, IL 60510, USA}

\author{M.~Schubnell}
\affiliation{Department of Physics, University of Michigan, Ann Arbor, MI 48109, USA}

\author{S.~Serrano}
\affiliation{Institut d'Estudis Espacials de Catalunya (IEEC), 08193 Barcelona, Spain}
\affiliation{Institute of Space Sciences (ICE, CSIC),  Campus UAB, Carrer de Can Magrans, s/n,  08193 Barcelona, Spain}

\author{I.~Sevilla-Noarbe}
\affiliation{Centro de Investigaciones Energ\'eticas, Medioambientales y Tecnol\'ogicas (CIEMAT), Madrid, Spain}

\author[0000-0002-3321-1432]{M.~Smith}
\affiliation{School of Physics and Astronomy, University of Southampton,  Southampton, SO17 1BJ, UK}

\author[0000-0001-6082-8529]{M.~Soares-Santos}
\affiliation{Brandeis University, Physics Department, 415 South Street, Waltham MA 02453}

\author[0000-0002-7822-0658]{F.~Sobreira}
\affiliation{Instituto de F\'isica Gleb Wataghin, Universidade Estadual de Campinas, 13083-859, Campinas, SP, Brazil}
\affiliation{Laborat\'orio Interinstitucional de e-Astronomia - LIneA, Rua Gal. Jos\'e Cristino 77, Rio de Janeiro, RJ - 20921-400, Brazil}

\author[0000-0002-7047-9358]{E.~Suchyta}
\affiliation{Computer Science and Mathematics Division, Oak Ridge National Laboratory, Oak Ridge, TN 37831}

\author{M.~E.~C.~Swanson}
\affiliation{National Center for Supercomputing Applications, 1205 West Clark St., Urbana, IL 61801, USA}

\author[0000-0003-1704-0781]{G.~Tarle}
\affiliation{Department of Physics, University of Michigan, Ann Arbor, MI 48109, USA}

\collaboration{(DES Collaboration)}

		\begin{abstract}

Transneptunian objects (TNOs) are a source of invaluable information to access the history and evolution of the outer solar system. However, observing these faint objects is a difficult task. As a consequence, important properties such as size and albedo are known for only a small fraction of them. Now, with the results from deep sky surveys and the Gaia space mission, a new exciting era is within reach as accurate predictions of stellar occultations by numerous distant small solar system bodies become available. From them, diameters with kilometer accuracies can be determined. Albedos, in turn, can be obtained from diameters and absolute magnitudes. We use observations from the Dark Energy Survey (DES) from November 2012 until February 2016, amounting to 4\,292\,847 CCD frames. We searched them for all known small solar system bodies and recovered a total of 202 TNOs and Centaurs, 63 of which have been discovered by the DES collaboration until the date of this writing. Their positions were determined using the Gaia Data Release 2 as reference and their orbits were refined. Stellar occultations were then predicted using these refined orbits plus stellar positions from Gaia. These predictions are maintained, and updated, in a dedicated web service. The techniques developed here are also part of an ambitious preparation to use the data from the Large Synoptic Survey Telescope (LSST), that expects to obtain accurate positions and multifilter photometry for tens of thousands of TNOs.

\end{abstract}

\preprint{DES-2017-xxxx}
\preprint{FERMILAB-PUB-xx-xxx-PPD}


		\section{Introduction} 
        \label{sec:intro}

The trans-neptunian region (30 AU distant from the Sun and beyond) is a world of small (diameters smaller than 2,400 km), faint (typically, $V>21$), and cold (20 -- 50 K) bodies. These are pristine objects, as well as collisional and dynamical remnants, of an evolved planetesimal disc of the outer solar system whose history and evolution can therefore be accessed from the trans-neptunian objects (TNOs).

Centaurs also play an important role in this study. They are located closer to the Sun in unstable orbits between Jupiter and Neptune, and it is generally accepted that they share a common origin with the TNOs. In this context, they serve as proxies to those more distant and fainter bodies \citep{2002AJ....123.1050F}.

Because of their large distances from the Sun, TNOs are difficult to observe and study. It is interesting to note that the 30-50 AU region is expected to contain 70,000 or more TNOs with diameters larger than 100 km \citep{2007MNRAS.375.1311I}. However, the Minor Planet Center\footnote{\url{https://minorplanetcenter.net/iau/mpc.html}} (MPC) lists, to date, a total of $\sim$2,700 TNOs/Centaurs and features like diameters, colors, taxonomy, presence of satellites are known for less than 15\% of these objects\footnote{\url{http://www.johnstonsarchive.net/astro/tnoslist.html}}. As a consequence, a number of questions about them, like their sizes, size distribution, and a relationship between size and magnitude, are poorly answered. The answers to these questions reveal the history of the trans-neptunian region and leads to the knowledge of its total mass \citep[see][for a comprehensive review and discussion of the trans-neptunian region]{2008ssbn.book.....B}.

A dramatic change in this scenario, however, is expected from the deep sky surveys. \citet{2009arXiv0912.0201L}, for instance, estimates that 40,000 TNOs will be observed by the Large Synoptic Survey Telescope (LSST) during its 10 years of operation. 

As far as the study of these objects through the stellar occultation technique is concerned, it is clear that the combination of large sky surveys and the astrometry from the Gaia space mission \citep{2018arXiv180409365G} will provide accurate occultation prediction for numerous bodies.

Although stellar occultations are transient events and still poorly predicted for most TNOs and Centaurs, it is the only ground-based technique from which sizes and shapes can be obtained with kilometer accuracies. Atmospheres can also be studied as their presence, or upper limits for their existence to the level of few nano-bars, can be inferred and modelled \citep[see][for details on sizes, shapes, and atmospheres from stellar occultations]{2009Icar..199..458W,2010Natur.465..897E,2011Natur.478..493S,2012Natur.491..566O,2013ApJ...773...26B,2015MNRAS.451.2295G,2016ApJ...819L..38S}. In addition, structures like rings \citep{2014Natur.508...72B,2017Natur.550..219O} or even topographic features \citep{2017AJ....154...22D} can be detected.

The Dark Energy Survey \citep[DES,][]{2005IJMPA..20.3121F} observations offer a considerable contribution to the study of small bodies in the solar system \citep[see][for an overview of the survey's capabilities]{2016MNRAS.460.1270D}. During its first three years of operations, 2013--2016, more than 4 million CCD images were acquired, where tens of thousands of solar system objects can be found. This considerable amount of data provides accurate positions and multi-filter photometry to, so far, more than one hundred TNOs and tens of Centaurs as faint as $r\sim$24.0. 

Here we present, from the above mentioned observations, positions, orbit refinement, and stellar occultation predictions for all known TNOs and Centaurs, 63 of the them discovered by the DES date-range for data as part of the tasks of its Transient and Moving Object (TMO) working group. One of these objects, 2014 UZ224, has already been studied in more detail from radiometric techniques by \citet{2017ApJ...839L..15G}.

In the next section, we briefly describe the DES. In Section 3, we describe the procedure to identify the known solar system objects in the images and the data reduction. In Section 4, we present results and data analysis. Conclusions and comments are presented in Section 5. Photometric data will be presented and explored in a separate paper.

	\section{The Dark Energy Survey} 
	\label{sec:des}

The DES is a survey that covers 5,000 square degrees in the $grizY$ bands of the southern celestial hemisphere. It aims primarily to study the nature of the dark energy, an unknown form of energy that leads to an accelerated expansion of the universe \citep[e.g.,][]{1998Natur.391...51P, 1998AJ....116.1009R, 2003RvMP...75..559P}. 

Observations within the survey are made with the Dark Energy Camera \citep[DECam,][]{2015AJ....150..150F}, a mosaic of 62 2k$\times$4k red-sensitive CCDs installed on the prime focus of the 4m Blanco telescope at the Cerro Tololo Inter-American Observatory (CTIO). The DECam has a field of view (FOV) of 3 square degrees and the wide-area survey images have, at a 10$\sigma$ detection level, a nominal limiting magnitude of $r=23.34$, with the final co-added depth being roughly one magnitude deeper \citep{2018arXiv180103177M}. The limiting magnitude is a quantity explained later in the text.

Considering only those observations made during the first three years of operation of the DES, the DECam acquired science images from more than 69,000 pointings or, more precisely, 4,292,847 individual CCD exposures in the five bands. This is an invaluable dataset to studies in several fields of astronomy \citep[see][]{2016MNRAS.460.1270D}, in particular, those related to transient events and moving objects.

	\section{Data and Tools} 
    \label{sec:data}

Our basic observational resource are the individual CCD images available from the DES database. In this database, the images taken until FEB/2016 were already corrected for a number of effects (crosstalk, bias, bad pixels, nonlinear pixel response, and flat field), in addition to image-specific corrections like bleed trails from saturated stars, streaks, and cosmic rays \citep[see][for a detailed description of the DES image processing pipeline]{2018arXiv180103177M}.

The set of tools used in this work are general, in the sense that they can be applied to any other survey or image database, and comprehensive, in the sense that they consider all necessary steps (in brief, identification of images with known Solar System bodies, astrometry, orbit refinement, and prediction of a stellar occultation).

These tools, described next, have been ingested in a high performance computational environment to form a pipeline in preparation to also use the data from the LSST. In fact, although LSST is expected to deliver astrometric accuracy ranging typically from 11 mas ({\it r}=21) to 74 mas ({\it r}=24) \citep{2009arXiv0912.0201L}, better astrometry (1-2 mas) is necessary to accurately predict stellar occultations by satellites of small bodies or grazing occultations by rings or by the main body itself, for example. Therefore, it is essential to have tools to independently determine accurate positions when needed.

	\subsection{Data retrieval and object search}

The very first step consists of obtaining the necessary information -- pointing, observing date, location in the DES database among others -- on all CCD images acquired during the first three years of observations within the DES. This was done through {\it easyaccess}\footnote{\url{https://github.com/mgckind/easyaccess}}, a friendly SQL-based tool to query the DES database. The result from such a query was a file containing the metadata from 4\,292\,847 CCD images. This file then feeds into the Sky Body Tracker \citep[SkyBoT,][]{2006ASPC..351..367B}.

SkyBoT is a project aimed at providing a virtual observatory tool useful to prepare and analyze observations of solar system objects. In addition to the web-interface service it offers, queries are also possible from the command line. The basic inputs to a cone search\footnote{A search based on a sky position and an angular distance from this position.}, for instance, are: IAU identification of the observatory, J2000 pointing coordinates of a given CCD image, observation date, and a region centered on the pointing coordinates. All these data come from the meta-data previously mentioned. The output is a text or VOTable file format with pieces of information on all known small solar system bodies inside the given region, such as their J2000 astrometric right ascensions and declinations, $V$ magnitudes, names and numbers (when they are numbered), and dynamical classes among others. Table~\ref{tab:skbt} lists the total number of TNOs and Centaurs found in the DES images as well as the expected number of objects for which positions can be determined from them. As we will see later in the text, these expected numbers (column 4 in particular) were surpassed.

The result of the search with the SkyBoT was a file having 1,708,335 entries, most of them of around 140,000 main belt asteroid objects in more than 1.5 million CCD images. These objects, in addition to few thousand members of other dynamical classes also found in the images, are being treated separately.

Note that the detection of a TNO or Centaur is not expected for all selected CCD images. Objects that are faint ($V\gtrsim 24.0$) in the DES images, or images taken under non transparent sky, may not provide a detectable signal of the target. The most frequent exposure time of the DES frames presented here is 90 seconds (see \citet{2018arXiv180103177M}). 

\begin{deluxetable}{l|cccc}
\tablecaption{Statistics of known TNOs and Centaurs in the DES images from the three first years of survey.
\label{tab:skbt}}
\tablehead{
\colhead{Dynamical} & \colhead{Total} & \colhead{Total} & \colhead{Expected} & \colhead{Expected} \\
\colhead{class\tablenotemark{a}} & \colhead{objects} & \colhead{observations} & \colhead{objects} & \colhead{observations}
}
\colnumbers
\startdata
TNOs      & 270 & 16,537  & 84 & 3,010 \\
Centaurs  & \  67 & \ 2,519  & 13 &\ \ 333 \\
\enddata
\tablecomments{\scriptsize Columns (2) and (3): Total number of TNOs, Centaurs, and their respective observations, as alerted by the SkyBoT among the observations made by the DES until FEB/2016.
Columns (4) and (5): Expected total number of TNOs, Centaurs, and their respective observations, under the following constraints: ($V\leq24.0$) and ephemeris uncertainty $\leq2^{\prime\prime}$ in both right ascension and declination. The visual magnitude as well as the positional uncertainties were also obtained from the SkyBoT.}
\tablenotetext{a}{\scriptsize As provided by the SkyBoT.}
\end{deluxetable}

		\subsection{Astrometry}

Our astrometric tool is the Platform for Reduction of Astronomical Images Automatically \citep[PRAIA,][]{2011gfun.conf...85A} package. PRAIA was conceived to determine photometry and accurate positions from large numbers of CCD images as unsupervised as possible. Its use and performance have been reported by various works \citep[see, for instance][]{2013MNRAS.430.2797A,2015A&A...583A..59T, 2016MNRAS.462.1351G} from reference frame to solar system studies. The reference catalog used here for astrometry is the Gaia Data Release 2 \citep{2018arXiv180409366L}. All differences in right ascension as well as all uncertainties related to measurements along right ascension are multiplied by the cosine of the declination.

A Intel(R) Xeon(R) CPU E5-2650 v4@2.20GHz configuration, using 40 cores, reduces 1,000 CCD images in 20 minutes from a parallelized run of PRAIA. A total of 12,561 CCD images were treated here. 

The presence of distortion effects, also known as Field Distortion Pattern (FDP), are expected in detectors with large FOVs such as that of the DECam. Common solutions are, e.g., the use of a high-degree polynomial (not always recommended) to relate CCD and gnomonic coordinates of reference stars, the brute-force determination of a distortion mask \citep[e.g.,][]{2010A&A...515A..32A}, and the construction of an empirical model that takes into consideration effects due to the atmosphere and the instrument. This last one was the solution adopted here to correct for the FDP.

Such a solution (hereafter C0) is based on the model developed by \citet{2017PASP..129g4503B} and was the first step toward the determination of positions. C0 provides corrections for the instrumental distortion effects including color terms from the optics. A first degree polynomial can be subsequently used to relate CCD and gnomonic coordinates of reference stars, providing reliable solutions from fields with low star densities. Observed positions will be sent to the MPC.

		\subsection{Orbits}

The refinement of orbits is obtained with the code Numerical Integration of the Motion of an Asteroid \citep[NIMA,][]{2015A&A...584A..96D}. NIMA starts from existing orbital parameters and then iteratively corrects the state vector from the differences between observations and computed positions through least squares. NIMA adopts a specific weighing scheme that takes into account the estimated precision of each position ($\sigma_i$), depending on the observatory and stellar catalog used as reference to determine the observed positions, the number of observations obtained during the same night in the same observatory ($N_i$) as well as a possible bias due to the observatory ($b_i$). The final variance of the observation $i$ is given by $\omega_i^2 = N_i b_i^2 +\sigma_i^2$. As a consequence, the weight is given by $1/\omega_i^2$. This weighing scheme is particularly relevant when we consider old epoch positions that do not use the Gaia catalog as reference. 

The values used in the NIMA weighing scheme are described in \cite{2015A&A...584A..96D} and were consolidated before the release of the astrometric data from the Gaia mission. Therefore, the code was improved to profit from the DES observations and from the Gaia releases. In this way, we have adopted $\sigma_i=b_i=0.125^{\prime\prime}$ for observations reduced with the Gaia DR1 and $\sigma_i=b_i=0.1^{\prime\prime}$ for observations reduced with the Gaia DR2. We emphasize that this latter is the case of DES observations presented here. 

It is possible to run NIMA, with the help of few scripts, in an unsupervised way so that it is suitable for a pipeline. One of its outputs is the object ephemeris in a format ({\it bsp} -- binary Spacecraft and Planet Kernel) that can be readily used by the SPICE/NAIF tools \citep{1996P&SS...44...65A,Act18} to derive the state vector of a given body at any time.

		\subsection{Prediction of stellar occultations}

The prediction of an occultation event is given by prediction maps that show where and when, on the Earth, such an event can be observed. This involves the knowledge of the Earth's position in space, the geocentric ephemeris of the occulting body, and a set of stellar positions in the neighborhoods of the sky path of the occulting object as seen by a geocentric observer \citep[see details in][]{2010A&A...515A..32A}. Note that, with the astrometry from Gaia, the uncertainties in predictions rest completely upon the accuracy of the ephemerides. 

A dedicated website, as presented in the next section, provides these occultations maps where many events occuring during daylight are also shown. This is done so that we are aware of even those ones that can be observed near the Earth terminator.

		\section{Results and analysis} 	
        \label{sec:results and analysis}

The high quality of the DES images provided us an accurate set of positions within the range of the observed magnitudes. As a consequence, the objects studied here were grouped according to the number of observations and the uncertainty of their existing ephemeris, rather than on the accuracy of the observed positions. Note that we use the ephemeris positions as a primary parameter to identify the observed position of a given TNO/Centaur in the images.

\subsection{Filtering}
\label{sec:filtering}

The determination of positions of TNOs and Centaurs from the DES images was subject to at least three constraints. The first one is that the ephemeris position of the target falls inside a box size of $4^{\prime\prime}\times 4^{\prime\prime}$ centered on its observational counterpart. The second is an iterative $3\sigma$ filtering on the offsets, as obtained from the differences between observations and a reference ephemeris, to eliminate outliers. The third constraint is based on a brief inspection of the magnitudes as obtained from the DES database for each filter. Differences larger than $\Delta=0.9$ magnitude between the brightest and faintest values in each filter, when multiple measurements were available, were investigated and eventually eliminated. This value of $\Delta$ takes into account a maximum variation of $\sigma_S=0.15$ (absolute value) in the magnitude due to the object's rotation, a maximum uncertainty of ${\sigma_M}=0.1$ in the observed magnitude, and a maximum variation of $\sigma_P=0.25$ (absolute value) in the observed magnitude due to the phase angle. In other words, $\Delta\sim 3\times\sqrt{{\sigma_S}^2+{\sigma_M}^2+{\sigma_P}^2}$.

These constraints were expected to provide a reliable identification of the solar system objects in the images with minimum elimination of good data. However, a preliminary orbit fitting of some objects still showed the presence of real outliers (misidentifications). To solve this, a fourth filter was applied to our data and affected mostly those sources whose ephemerides presented large uncertainties (extension and doubtful sources, see Section~\ref{sec:organization}). Basically, this filter looks for positional offsets (observed minus ephemeris positions) that are clustered and, from them, a mean ($x_0$) and standard deviation ($\sigma_0$) are determined. Then, any offset that differs from $x_0$ by more than $N\times\sigma_0$ is eliminated. Most frequently, $N=5$ was used. This filter is more resistant to the presence of outliers than the 3$\sigma$ filtering mentioned above.

As a result from this process, misidentifications of TNOs and Centaurs from the images were reduced to a minimum, although real outliers may still be found mostly among the doubtful sources with large ephemeris uncertainties.

\subsection{Organization}
\label{sec:organization}

\begin{figure*}
\centerline{\includegraphics[scale=0.6]{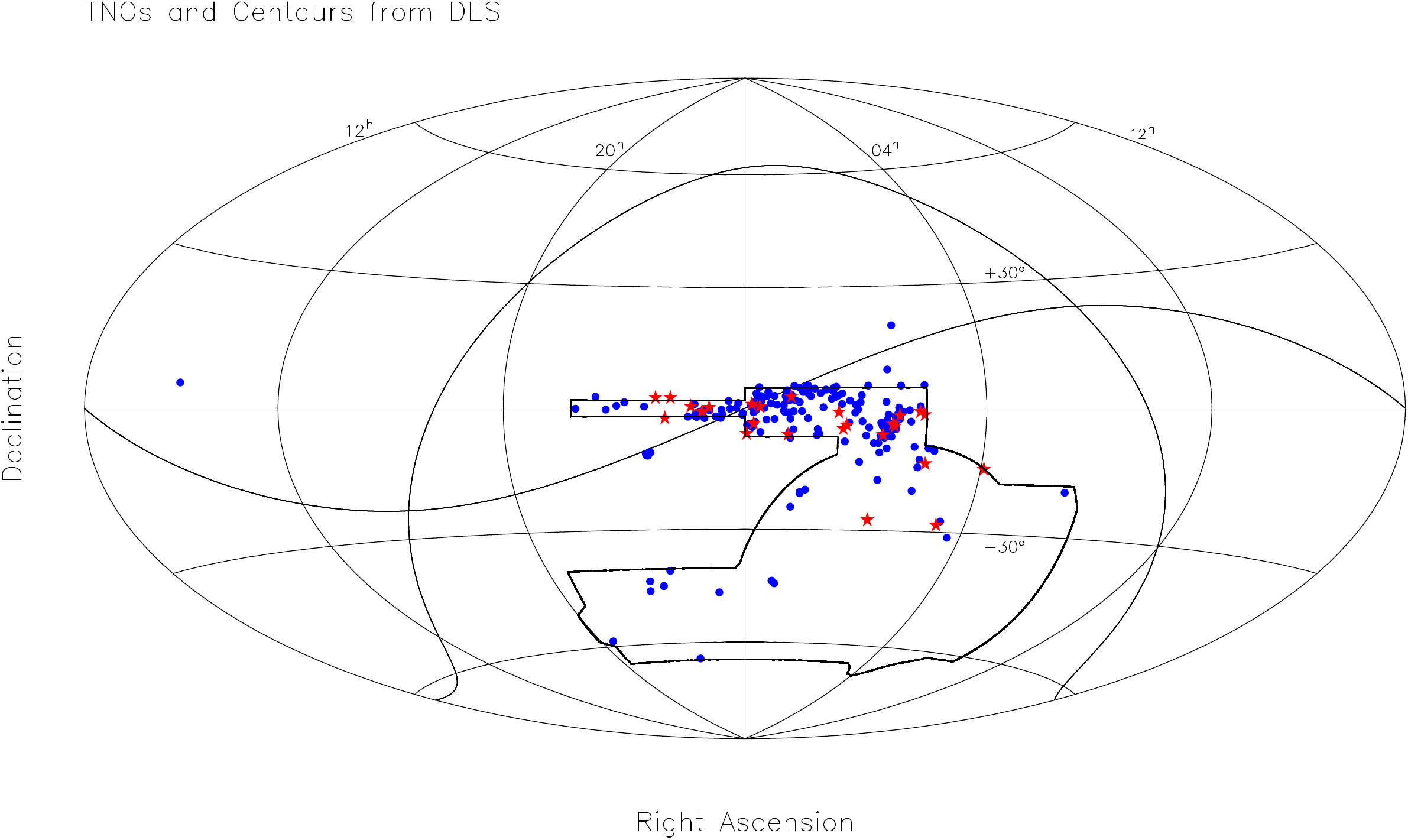}}
\caption{Hammer-Aitoff equal-area projection of the sphere for the TNOs (blue dots) and Centaurs (red stars) for which a position was determined. The ecliptic and Galactic planes, as well as the DES footprint, are also represented by black lines. Some fields are clearly outside the DES footprint. They refer to observations associated to the Vimos VLT deep survey \citep[leftmost blue dot -][]{2005A&A...439..845L}, to the LIGO event G211117 \citep[the two northermost blue dots -][]{2016ApJ...826L..29C}, and to DES engineering time (blue dots close to the ecliptic, at right ascension $\sim$22.4h).
\label{fig:sky}}
\end{figure*}

Our results in astrometry are organized in Tables \ref{tab:selkbores} to \ref{tab:kbodoubt} (Appendix A), and the respective source distribution in the sky can be seen in Fig.~\ref{fig:sky}. 

Table~\ref{tab:selkbores} (main) consider those sources for which the $1{\sigma}$ ephemeris uncertainty ($\sigma_{E}$) in RA/DEC is smaller than or equal to $2^{\prime\prime}$ for TNOs and Centaurs and the number of observations (\textmd{N}) is greater than or equal 3. Table~\ref{tab:kboextend} (extension) consider those sources for which the ephemeris uncertainty is $2^{\prime\prime}$$<$$\sigma_{E}$$\leq$$12^{\prime\prime}$ and $\textmd{N} \ge 5$. Table~\ref{tab:kbodoubt} (doubtful) consider the remaining sources. All the ephemeris uncertainties used in these tables were obtained from the JPL on 27 April 2018 and are referred to 2014-Jan-01 at 0h UTC. Note that these uncertainties are given as they appear in their respective ephemerides, that is, $3\sigma$ values.

\subsubsection{The extension table: rationale}

Most (90\%) of the CCD images treated here have less than 1,100 sources. Knowing that the size of one CCD in the DECam is $\sim 9^{\prime}\times18^{\prime}$, we can consider that there is 1 field object\footnote{Any signal on the CCD that is recognized as an object (star, solar system object, etc).}, on average, inside a box of $24^{\prime\prime}\times 24^{\prime\prime}$. In this way, it is expected that a box of this size centered on the ephemeris (calculated) position of an object in Table~\ref{tab:kboextend} contains the respective observed position and a field star. If any of them falls inside a box of $4^{\prime\prime}\times 4^{\prime\prime}$ around the ephemeris position, then this observed position is flagged as an eligible target. If not eliminated by the other steps of the filtering process, then this observed position is selected to refine the respective orbit.

\begin{figure*}
\includegraphics[width=.48\textwidth]{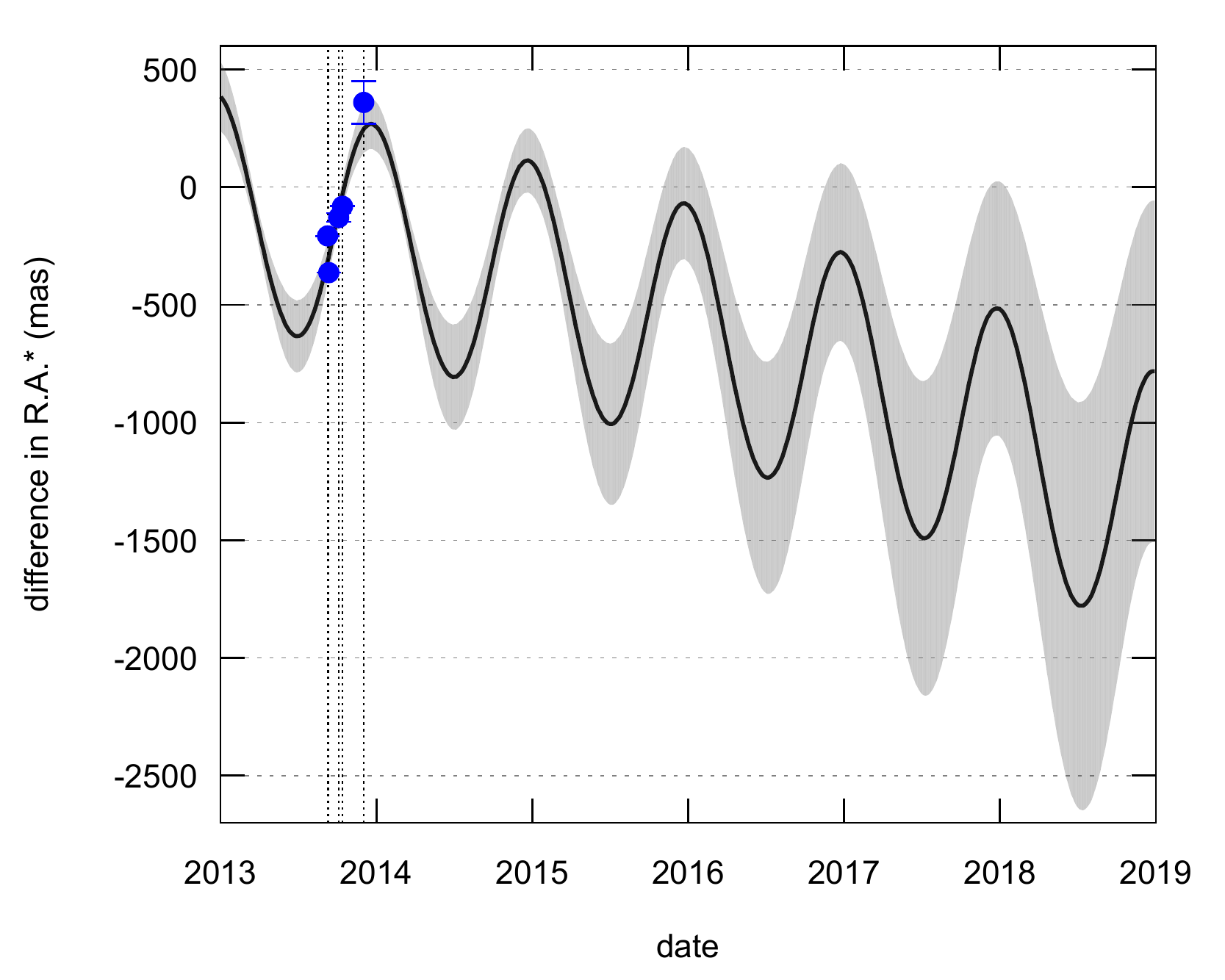} \hfill \includegraphics[width=.48\textwidth]{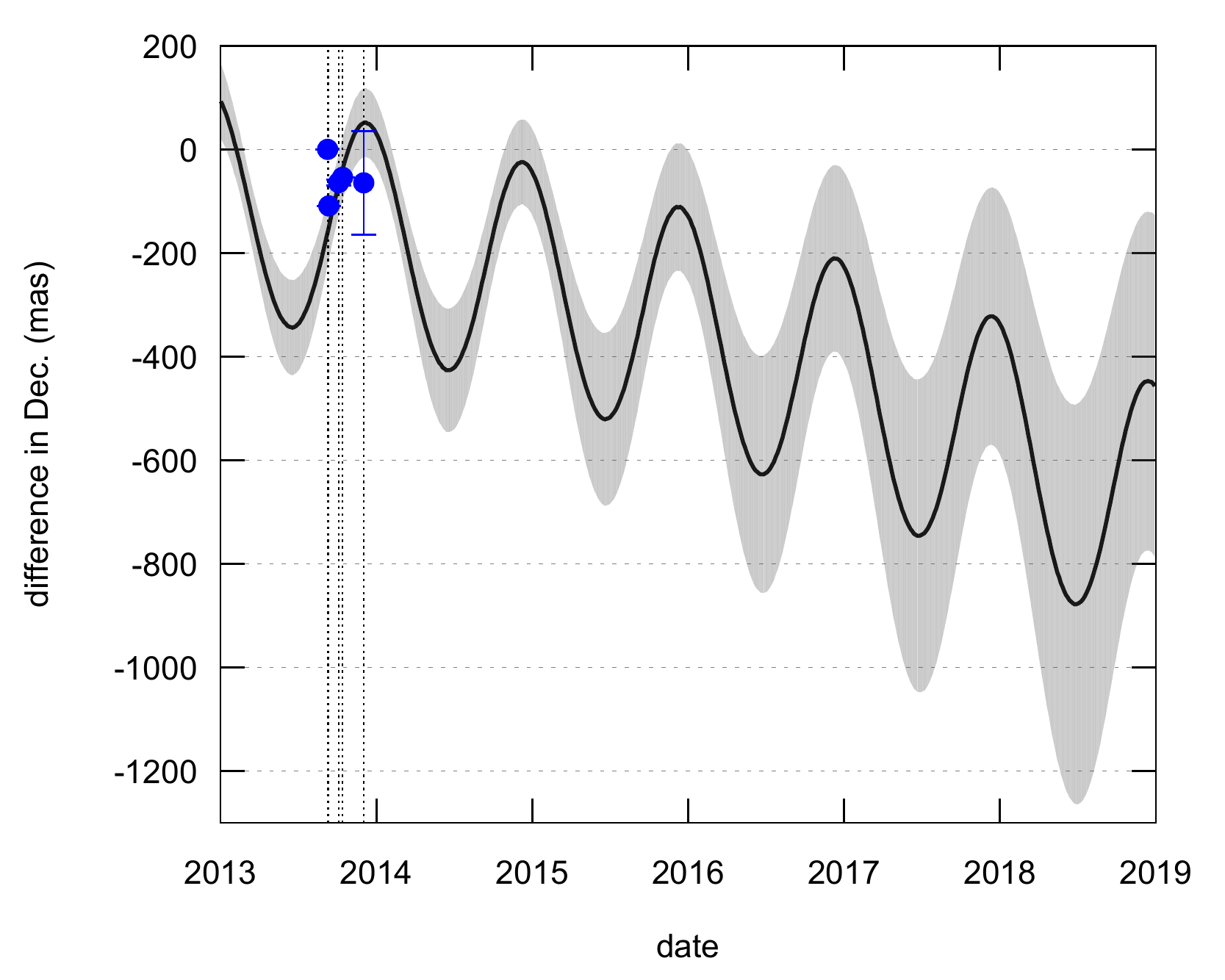}
\caption{Difference (black lines) in right ascension (left panel) and declination (right panel) between the orbit determined with NIMA and that from the JPL (version: JPL\#4) for the TNO 2002 PD149. In the same way, blue dots are the differences between the observed positions and those from the JPL ephemeris. This object belongs to the TNO extension group, Table~\ref{tab:kboextend}. The sense of the differences is NIMA minus JPL.
\label{fig:occmaps2}}
\end{figure*}

We adopted the number 5 as the minimum number of filtered (see Section~\ref{sec:filtering}) selected positions that an object with ephemeris uncertainty $2^{\prime\prime}$$<$$\sigma_{E}$$\leq$$12^{\prime\prime}$ must have to appear in the extension table. Orbits for the objects in this table do not have the same quality as those for objects in Table~\ref{tab:selkbores}. However, as illustrated by Fig.~\ref{fig:occmaps2} (compare it to Fig.~\ref{fig:occmaps1}, panel (a), shown later in the text), the 5 or more positions of each object in that table are a relevant contribution to the refinement of their respective orbits.

\subsection{Accuracies}
\label{sec:accuracies}

In the astrometric analysis of these images, it is interesting to introduce here the concept of {\it limiting magnitude}, as presented by \citet{Nei15}, and also discussed by \citet{2018arXiv180103177M}. 

\begin{figure*}
\gridline{\fig{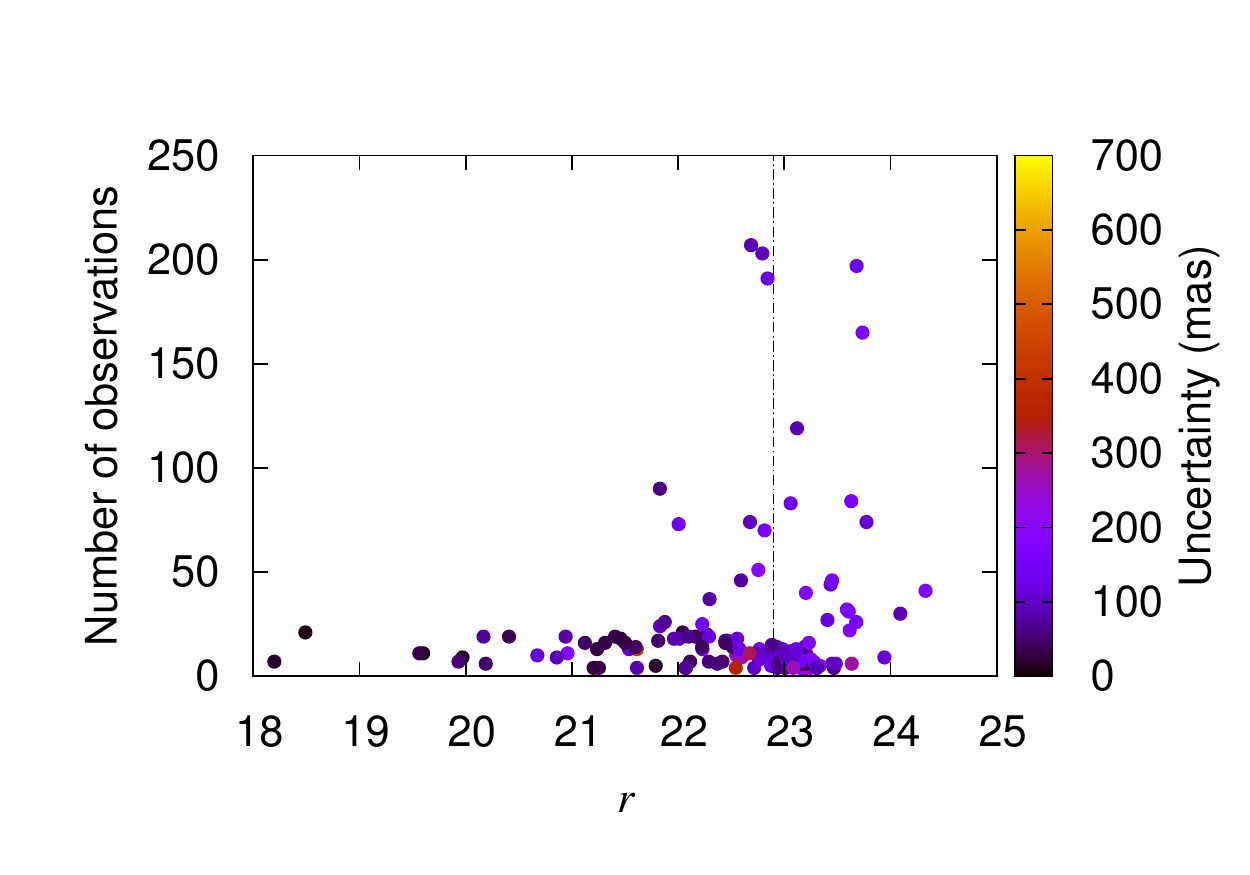}{0.5\textwidth}{}
          \fig{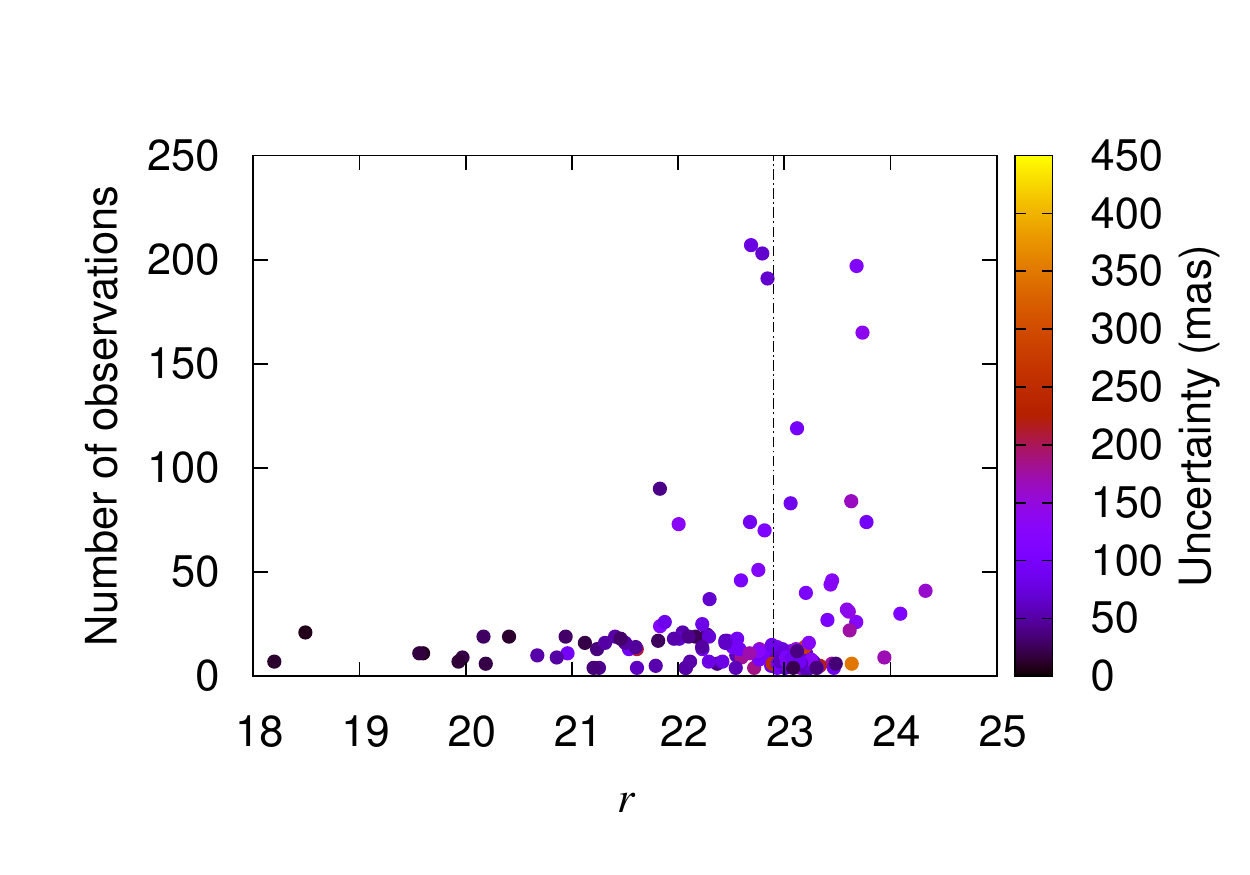}{0.5\textwidth}{}
         }
\gridline{\fig{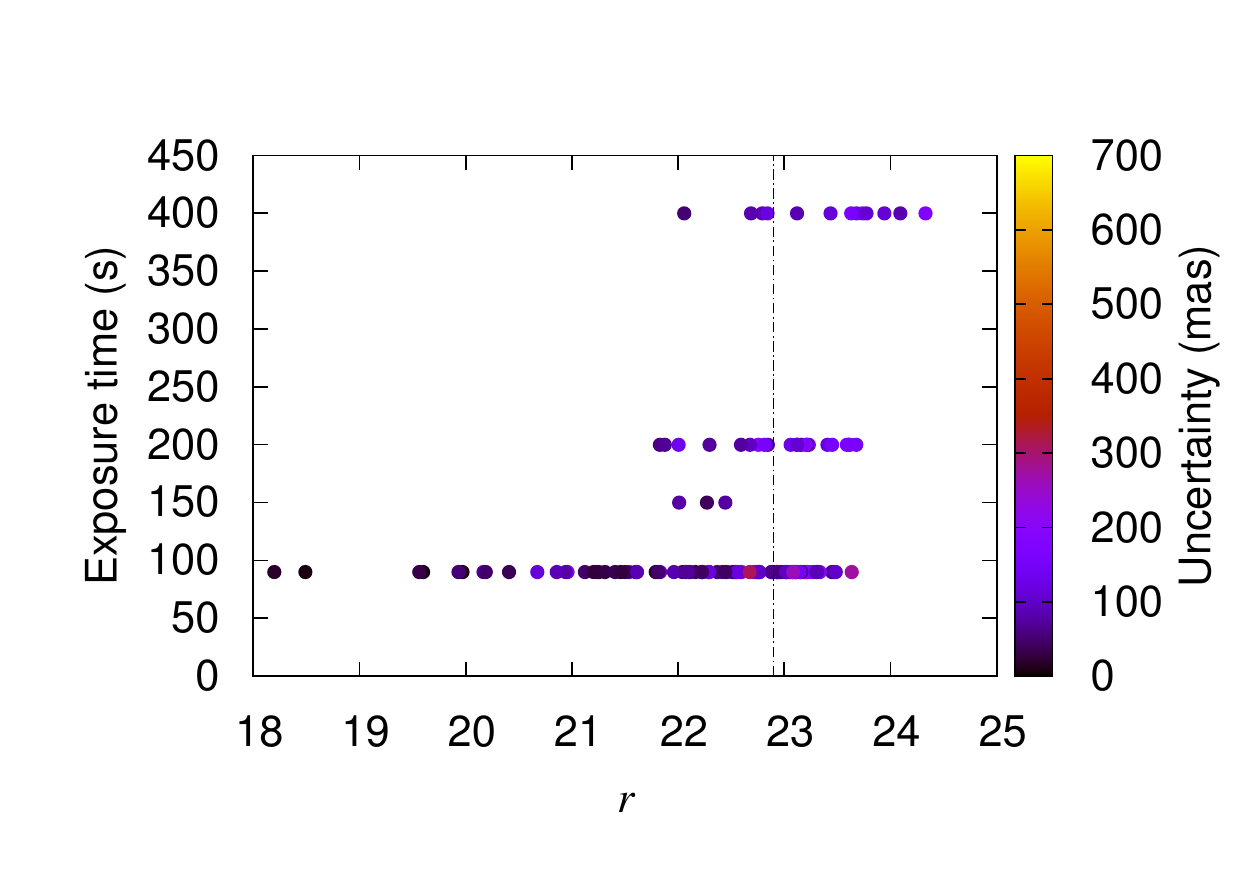}{0.5\textwidth}{}
          \fig{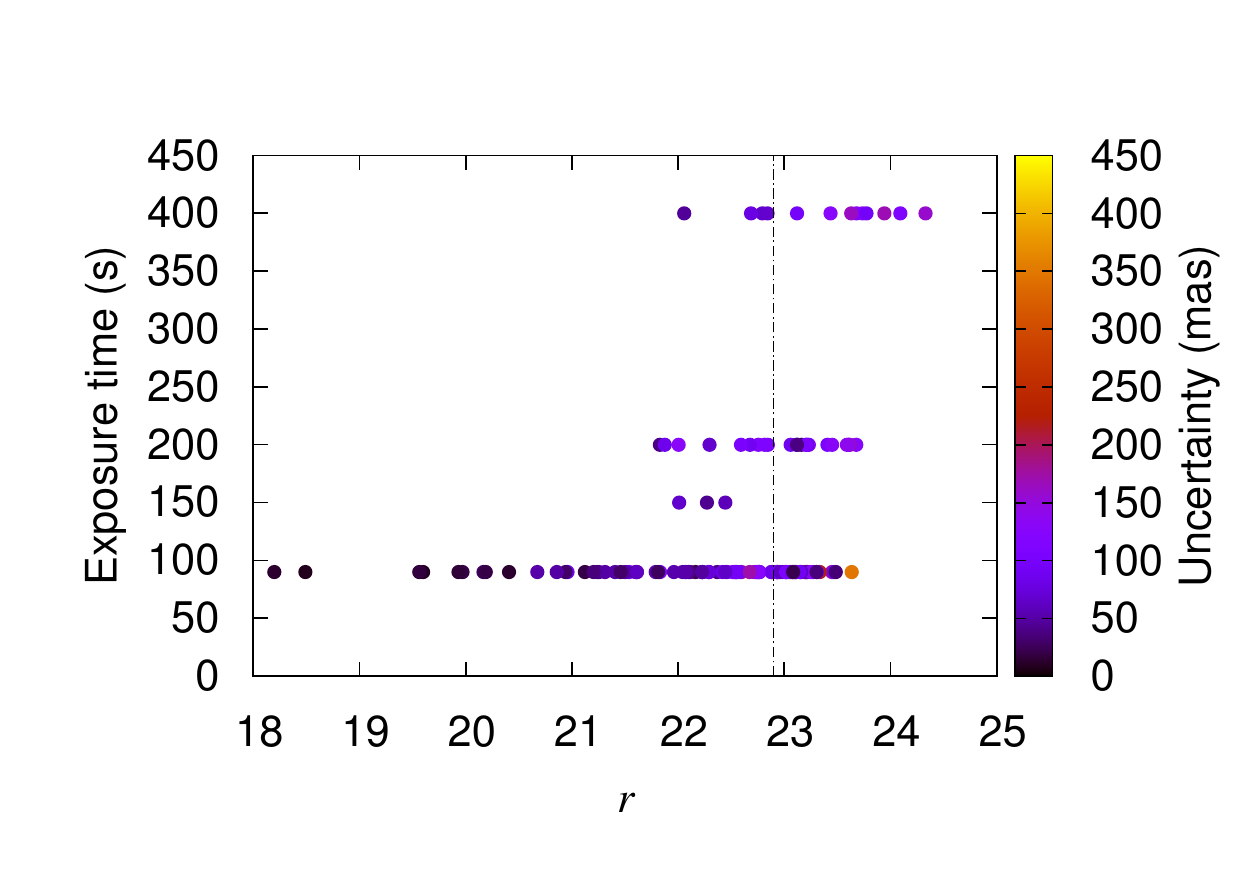}{0.5\textwidth}{}
         }
\caption{Positional uncertainty as a function of the magnitude and the number of observations in right ascension (left panels) and declination (right panels) for the TNOs and Centaurs in Tables from~\ref{tab:selkbores} to \ref{tab:kboextend}. In the upper panels, the number of observations is given as a function of the magnitude. In the lower panels, the exposure times are given as a function of the magnitude. In case of different exposure times for the same object, the longest one was considered. In all panels, the positional uncertainty is given in mas and are color coded. The dashed line gives the median value (22.9) of the limiting magnitude in the $r$-band for these observations. In the upper panels, the TNO (437360) 2013 TV158 (see Table~\ref{tab:selkbores}) is not shown due to its large number of observations (438).}
\label{fig:errxmagsnobs}
\end{figure*}

The limiting magnitude is that at which the magnitude of a star is measured with an uncertainty of 0.1 mag. It can be shown to be related to a quantity $\tau$ by
\begin{equation}
m_{lim}=m_{0}+1.25\log\tau,
\end{equation}
where $\tau$ is a scaling factor to the actual exposure time (given by the image header). As a consequence, an {\it effective exposure time} can be defined as $\tau~\times$ nominal exposure time. The $\tau$ quantity and the limiting magnitude, therefore, can be used as a quality parameter for a given image. In order to determine the limiting magnitude in the {\it r}-band shown in Figs.~\ref{fig:errxmagsnobs} and \ref{fig:efficiency}, the value $m_{0}=23.1$ was taken from \citet{Nei15} and the values of $\tau$ were obtained directly from the DES database for each CCD \citep{2018arXiv180103177M}.

The accuracy of the observations for the objects presented in Tables \ref{tab:selkbores} to \ref{tab:kboextend} (columns 5 and 6) is illustrated by Fig.~\ref{fig:errxmagsnobs}, where the average limiting magnitude (22.9) in the $r$-band (dashed line) sets a rough limit in the upper panels from which the uncertainties become larger, mainly when the number of observations is low. It also shows that the sources with a large number (hundreds) of observations have magnitudes that are close to or fainter than this limiting magnitude.

Two relevant features are shown by Fig.~\ref{fig:errxmagsnobs}. Firstly, the lower panels show that, even in frames with the shortest exposure time (90 s), we detect sources with {\it r} as faint as $\sim$ 24.0 with a quality that is comparable to those from frames with an exposure time of 400 s thanks to the excellent quality of the images. It is worth mentioning that the faintest objects are more than 1 magnitude fainter than the average limiting magnitude in the $r$-band. Secondly, it is also possible to note that the range of uncertainties in right ascension is wider than that in declination. This feature most probably results from the fact that the ephemeris uncertainties (columns 3 and 4, Tables~\ref{tab:selkbores} to \ref{tab:kbodoubt}) are, on average, larger in right ascension than in declination, since we do not verify such a large difference between our measurements in RA and DEC as discussed below.

The standard deviations in Tables~\ref{tab:selkbores} to \ref{tab:kbodoubt} (columns 5 and 6), obtained from the differences between the observed positions and those from the respective JPL ephemeris, is a common way to express the positional accuracy of solar system targets. These differences vary as a function of time so that, in the present study, the standard deviations provided by these columns numerically overestimate the internal accuracy (or repeatability) of the astrometric measurements.

A second astrometric empirical model (hereafter C1), also developed by the DES collaboration and based on \citet{2017PASP..129g4503B}, provides improved astrometric solutions for all of the good-quality wide-survey DES exposures for years 1 through 4 of the survey. From C1, instrumental solutions are believed accurate to smaller than 3 mas RMS per coordinate. As a consequence, every DES astrometric measurement will be limited by the stochastic atmospheric distortions, typically $\sim$10 mas RMS in a single exposure within this solution. Note that, as compared to C0, C1 is available to a smaller set of DES exposures. 

We compared the positions we determined for TNOs and Centaurs to all those ones resulting from C1. This comparison is summarized in Table~\ref{tab:comparison}, where all differences we found between our results and those from C1 were kept. It is important to note, however, that C1 does not provide a solution for all CCDs. We stress that C1 is only used to provide a more realistic estimate of the internal accuracy of our measurements as well as a comparison between our positions and those from the most recent astrometric empirical model developed by the DES collaboration. C1 does not participate in any of the astrometric determinations provided here.

\begin{deluxetable}{l|ccccc}
\tablecaption{Differences between the astrometric results presented here and the DES empirical model
\label{tab:comparison}}
\tablehead{
\colhead{Type} & \colhead{$\Delta\alpha{\rm cos}\delta$} & \colhead{$\Delta\delta$} & \colhead{$\sigma_{\alpha}{\rm cos}\delta$} & \colhead{$\sigma_{\delta}$} & \colhead{Measurements} \\
 & \multicolumn2c{(mas)} & \multicolumn2c{(mas)} &
}
\colnumbers
\startdata
TNO      &  3  &  $-$4 & 11  & 9  & 142\\
Centaur  &  2  &  $-$5 & 12  & 5  &  22\\
\enddata

\tablecomments{Columns (2) and (3): average of the differences between this work and the empirical model in right ascension and declination, respectively. Columns (4) and (5): standard deviation from the measurements used to determine the values in Columns (2) and (3), respectively. Sense of the differences: this work minus empirical model.}
\end{deluxetable}

The standard deviations shown in Table~\ref{tab:comparison} (columns 4 and 5) are a more reliable estimate of the internal accuracy of our measurements, as compared to those obtained in Tables \ref{tab:selkbores} to \ref{tab:kboextend}. This internal accuracy is given by the standard deviation of the measurements, not of the mean. Therefore, the small systematics between both solutions (columns 2 and 3) can not be considered negligible. Part of them, at least, may be explained by the fact that the empirical model is based on the Gaia Data Release 1 \citep[Gaia DR1,][]{2016A&A...595A...4L}. It is also worth mentioning that, when our positions are referred to the Gaia DR1 (that is, the Gaia DR1 is used as reference for astrometry), the values of these standard deviations in right ascension and declination are more similar to each other.

On the other hand, a realistic estimate of the final positional accuracy of the targets (or how accurate their equatorial coordinates are given in the International Celestial Reference Frame \citep[ICRF,][]{1998AJ....116..516M}) can be obtained from the root mean square (RMS) of the reference stars, as given by the differences between their observed and catalog positions, and the precision in the determination of the object's centroid. This latter as well as the RMS of the reference stars for different filters and magnitude ranges are provided by Table~\ref{tab:sigref}. In this context, this final accuracy to both equatorial coordinates is obtained, at the $1\sigma$ level, from the quantity
\begin{equation}
\sigma_F=\sqrt{{\sigma_C}^2+{\sigma_R}^2},
\end{equation}
where $\sigma_C$ is the uncertainty in the determination of the objects' centroid and $\sigma_R$ is the RMS of the reference stars. For the $r$ filter, for instance, 12 mas $< \sigma_F < $ 20 mas. 

\begin{deluxetable}{l|cccc|cccc}
\tablecaption{Overall uncertainty values
\label{tab:sigref}}
\tablehead{
\colhead{Mag. interval} & \colhead{g} & \colhead{r} & \colhead{i} & \colhead{z} & \colhead{g} & \colhead{r} & \colhead{i} & \colhead{z} \\
 & \multicolumn4c{centroid (mas)} & \multicolumn4c{reference stars (mas)}
 }
\colnumbers
\startdata
$18\leq $mag$<$ 19  &   7 &   5 &   5 &   5 &  14 &  11 &  11 &  10   \\
$19\leq $mag$<$ 20  &  11 &   6 &   5 &   6 &  14 &  12 &  11 &  10   \\
$20\leq $mag$<$ 21  &  17 &   9 &   7 &   8 &  15 &  12 &  11 &  11   \\
$22\leq $mag        &  26 &  13 &  10 &  12 &  15 &  15 &  12 &  12   \\
\enddata
\tablecomments{Column1 (1): magnitude interval. Columns (2) to (5): precision in the determination of the centroid of TNOs and Centaurs as a function of the magnitude in a given filter. Columns (6) to (9): RMS of the reference stars as a function of the magnitude in a given filter. Note: these magnitudes do not correlate directly to those from Gaia.}
\end{deluxetable}

\subsection {Timing}

When dealing with solar system objects, the mid exposure time (time of the shutter opening plus half of the exposure time) is of particular importance. DECam has a shutter that takes a while (about 1s) to cross the focal plane, so the actual mean of the exposed time depends on the position in the focal plane. To compensate for this feature, the mid exposure time was obtained by adding
\begin{equation}
0.5\times({\rm exposure~time}+1.05{\rm s})
\end{equation}
to the value of the MJD as read from the image headers \citep[see][]{2015AJ....150..150F}. This becomes particularly relevant when dealing with objects in the inner solar system.

\subsection{Detection efficiency}

\begin{figure}
\centerline{\includegraphics[scale=0.65]{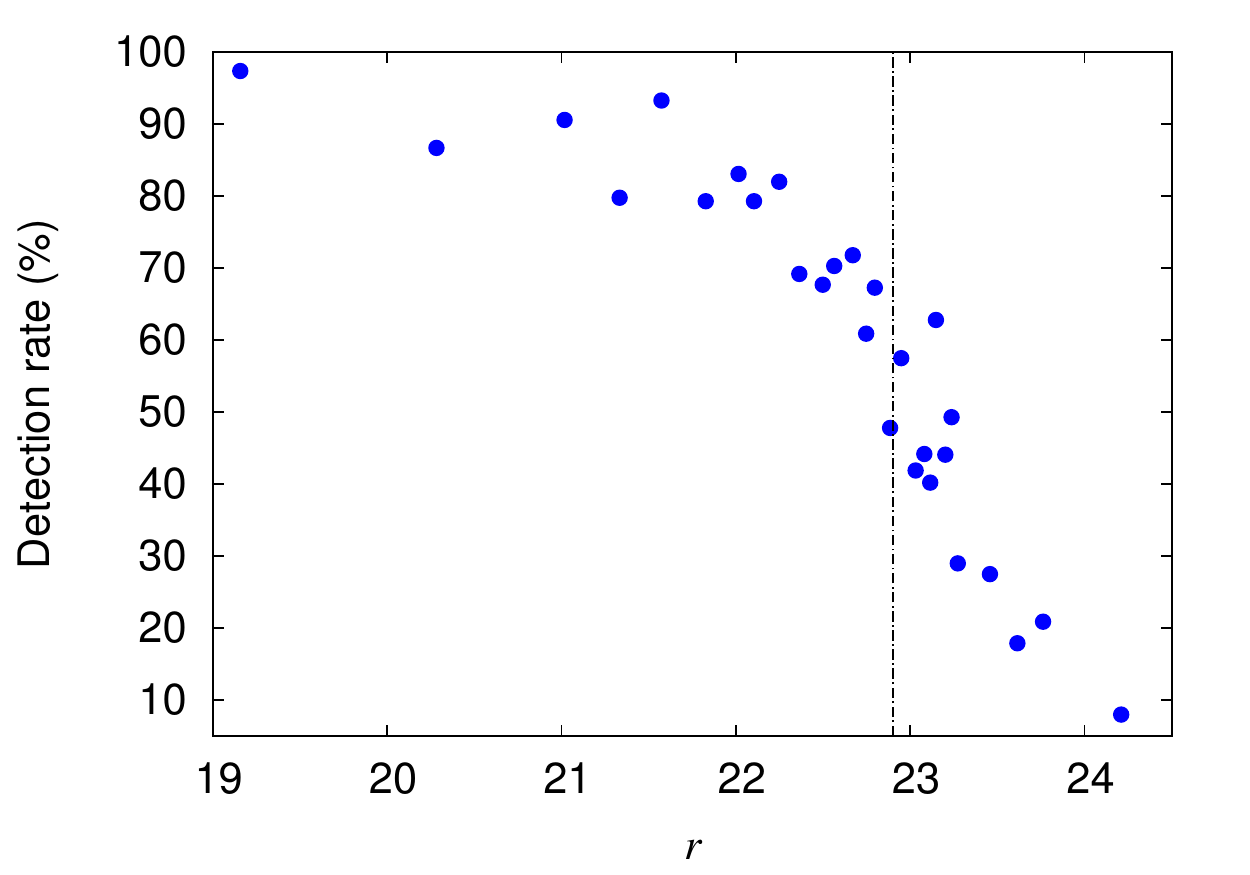}}
\caption{Detection efficiency as a function of the magnitude. No constraints on image quality are applied. The median limiting magnitude in the $r$-band (22.9), when accounting for $\tau$ is indicated by the vertical dashed line. Only TNOs and Centaurs in Tables~\ref{tab:selkbores} to \ref{tab:kboextend} with at least one measured magnitude in the {\it r} band were considered. 
\label{fig:efficiency}}
\end{figure}

In Fig. \ref{fig:efficiency} we show the detection efficiency as measured by the number of observed positions divided by the number of images for a given object. This figure has contributions from all images matched to objects in Tables \ref{tab:selkbores} and \ref{tab:kboextend}, including those taken under non photometric sky. This efficiency justifies the more favorable detection statistics shown in Table~\ref{tab:general} (column 3) as compared to the initial estimates given by Table~\ref{tab:skbt}. It is true that this latter, as opposed to Table~\ref{tab:general}, considers only those objects for which the uncertainty in the ephemeris is $\leq2^{\prime\prime}$. However, Table~\ref{tab:selkbores} alone, with 114 entries, corroborates this better performance.

\begin{deluxetable}{l|cccccccc}
\tablecaption{General numbers from images containing known TNOs and Centaurs
\label{tab:general}}
\tablehead{
\colhead{Type} & \colhead{Total} & \colhead{Ast} & \colhead{Pos} & \colhead{{\it g}} & \colhead{{\it r}} & \colhead{{\it i}} & \colhead{{\it z}} & \colhead{{\it griz}} 
}
\colnumbers
\startdata
TNO      &  270  & 177 & 3454  & 54 & 93 & 75 & 48 & 34\\
Centaur  &   67  &  25 &  545  &  9 &  6 &  9 &  6 &  3\\
\enddata
\tablecomments{Columns (2): total number of objects at start. Column (3): total number of objects with at least one position determined. Column (4): total number of positions determined. Columns (5) to (8): number of objects with at least 3 magnitude measurements in each indicated filter. Columns (9): number of objects with at least 3 magnitude measurements in each the four filters. Note: there were 4 positions measured in the {\it Y} band and none measured in the {\it u} band.}
\end{deluxetable}

\subsection{Orbits}

Orbit refinement is a straight forward process with NIMA, once positions are obtained. One ephemeris ({\it bsp} format) file is provided for each of the 177 TNOs and each of the 25 Centaurs (see Table~\ref{tab:general}), from which the J2000 equatorial heliocentric state vector of each body at any time\footnote{Limited to an interval of few decades (for instance, 2015 to 2025) to avoid large files.} can be obtained with the help of the SPICE/NAIF tools. 

As far as stellar occultations are concerned, it is enough to be aware of an occultation event one or two years in advance so that the object's ephemeris can be more intensively refined, if necessary, and the respective observation missions for the occultation can be organized. In this way, these ephemerides should be sufficiently accurate for 1-2 years after the most recent observations and constant updates must be provided. Ideally, we consider an ephemeris to be sufficiently accurate when its $1\sigma$ uncertainty is smaller than the angular size of the respective occulting body and very few objects $-$ (10199) Chariklo and Pluto among them $-$ profit from such ephemerides. Observations like those from the DECam are invaluable to change this scenario.

One disadvantage of the {\it bsp} files is that they do not carry information on uncertainties. Our dedicated website provides an orbit quality table in which uncertainties are given in steps of 6 months to each target. These uncertainties vary from few to hundreds of mas, depending mainly on the astrometric quality of the current epoch of observations.

The result of an ephemeris refinement is illustrated by Figs.~\ref{fig:occmaps2} (object from the Table~\ref{tab:kboextend}) and \ref{fig:occmaps1}, panel (a) (object from Table~\ref{tab:selkbores}). They compare the refined orbit with its counterpart from the JPL and show the uncertainty of the refined orbit along with the recently observed positions of the respective solar system body. Among others, it helps to have a first idea of the work still needed to reach suitable uncertainties for successful predictions.

The waving pattern seen in Fig.~\ref{fig:occmaps1}, panel (a), is a common feature. It is a consequence of the different heliocentric distances of the solar system bodies as determined from NIMA and the JPL combined with the Earth's motion around the Sun. Deep sky surveys like the DES also play a relevant role to improve the determination of these distances by providing observations at different phase angles. 

Orbits determined in this work can be found from the address \url{http://lesia.obspm.fr/lucky-star/des/nima}. To each object, a text file lists the positions here determined as well as the respective observational history from AstDys\footnote{\url{http://hamilton.dm.unipi.it/astdys/}} (MPC, if the object is not found in the AstDys) that were used to determine the orbit. The $1\sigma$ orbit uncertainty ($\sigma_{\alpha}\cos\delta$ and $\sigma_{\delta}$) is given for a period of two years in steps of six months from the last observation. Orbits themselves are available in {\it bsp} format. Details on the pages content are provided in a README file.

\subsection{The $a\times e$ plane}

\begin{figure*}
\centerline{\includegraphics[scale=1]{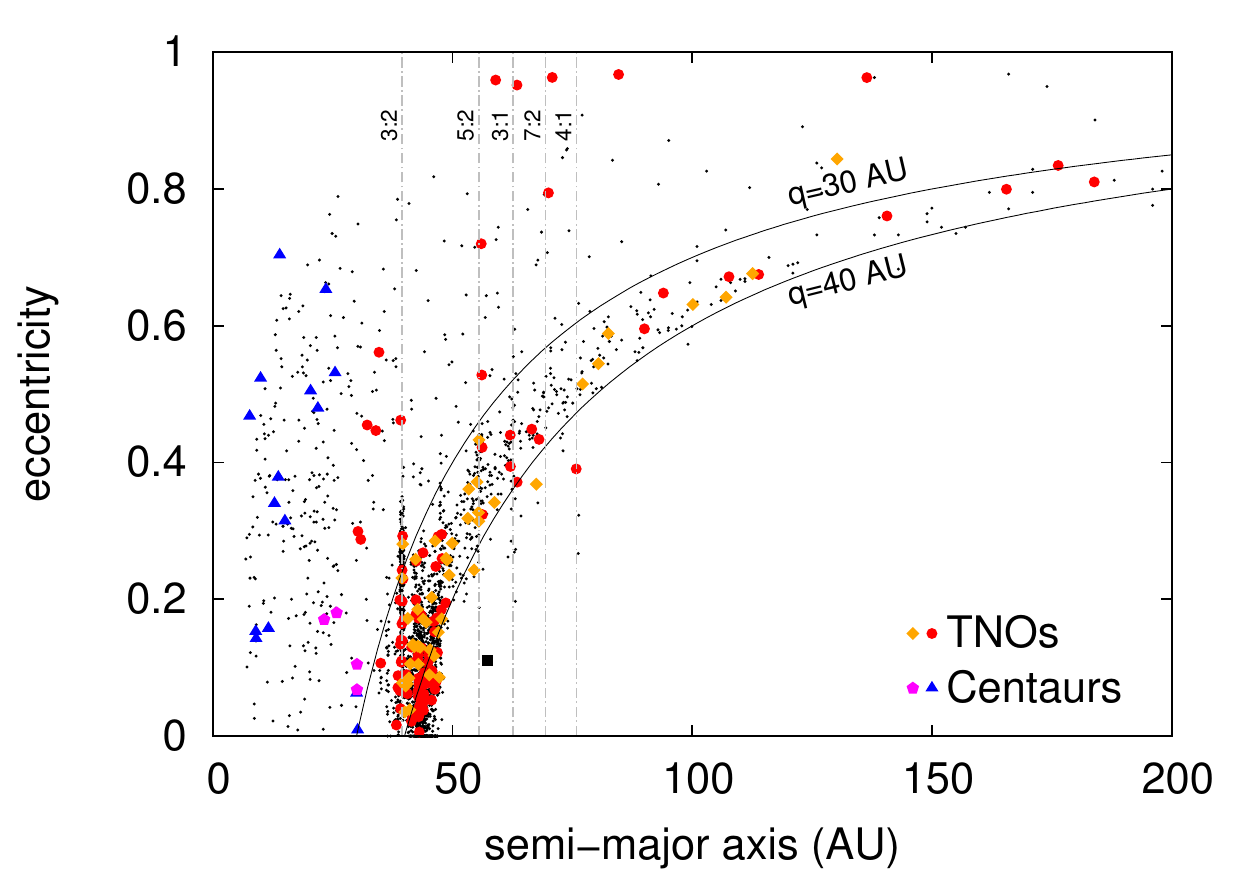}}
  \caption{Distribution of the TNOs and Centaurs whose orbits were refined (red circles, orange diamonds, blue triangles and magenta pentagons), along with others taken from the MPC (small back dots), in the $a \times e$ plane. Some mean motion resonances (MMR) with Neptune are also indicated. Objects discovered by the DES are given by orange diamonds (Centaurs) and magenta pentagons (TNOs). The black square shows the scattered disk object 2004 XR190, not observed by the DES.}
  \label{fig:axe} 
\end{figure*}

One important feature of surveys like DES is the possibility to provide a better insight on dynamical theories as the number of objects on which such theories may be employable increase through new discoveries. This is illustrated with the help of Fig.~\ref{fig:axe}.

Considering explicitly the osculating elements, it is interesting to note that the MPC lists, to date, 48 objects with $q>40$ AU and $a>50$ AU. They constitute a conspicuous population of detached objects, for which mechanisms capable of increasing their perihelia is a subject of interest. Three of these -- 2013 VD24, 2014 QR441, and 2005 TB190 -- were observed by the DES, the first two being discovered by the survey. All of them are shown in Fig.~\ref{fig:axe}.

\citet{2011Icar..215..661G} showed that there is a path between a scattering particle, induced by the migration of the giant planets, and the stable orbit similar to that of 2004 XR190 (black square in Fig.~\ref{fig:axe}, object not observed by the DES). This path results from a combination of Neptune's migration and mean motion resonance (MMR) plus Kozai resonance. One of the features of this dynamical path is that the new stable orbits escape the Neptune's MMR. The discovery of more objects by deep sky surveys with $q>40$ AU and $a>50$ AU may help to confirm this dynamical path.  

2013 VD24 (close to the 5:2 resonance) and 2014 QR441 (close to the 7:2 resonance) are potentially among these objects. Numerical integrations of the equations of motion are necessary to check if they are not trapped in the resonances indicated in Fig.~\ref{fig:axe}. A more detailed study is ongoing.

		\subsection{Occultation maps}

A dedicated website also provides access to occultation prediction maps for the TNOs and Centaurs in this work. 

\begin{figure*}[!h]
\gridline{\fig{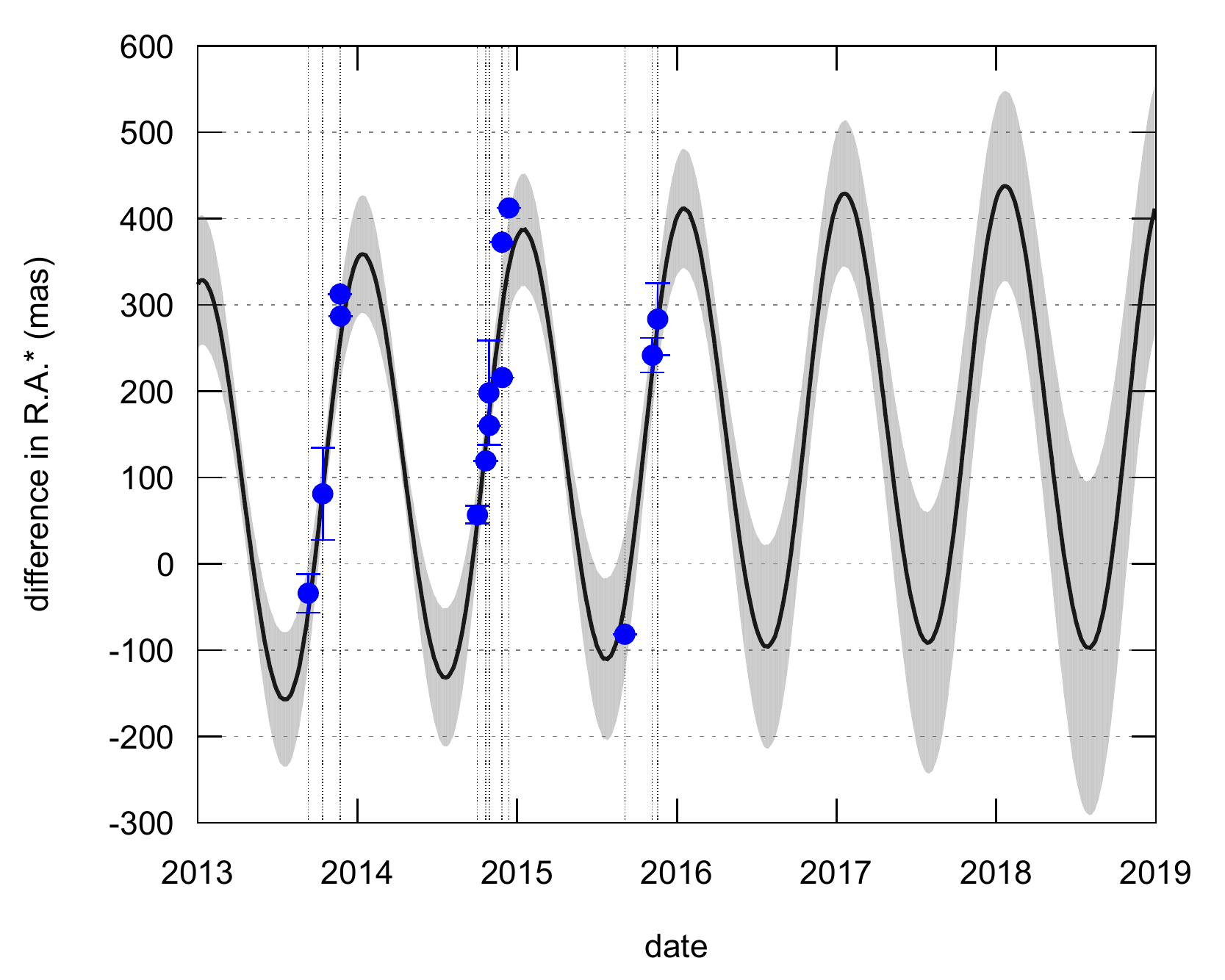}{.46\textwidth}{\hfill (a)} \fig{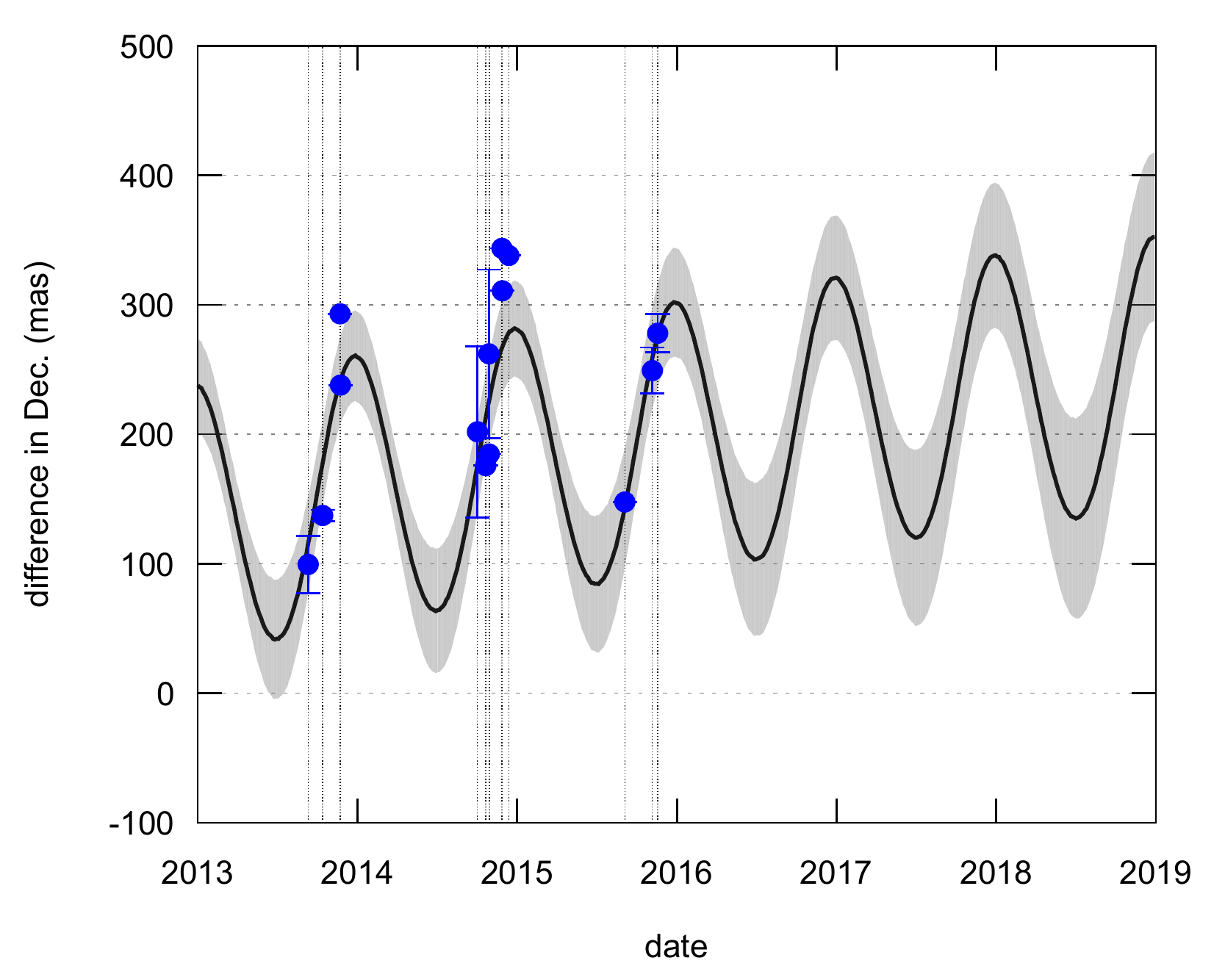}{.46\textwidth}{}}

\gridline{\fig{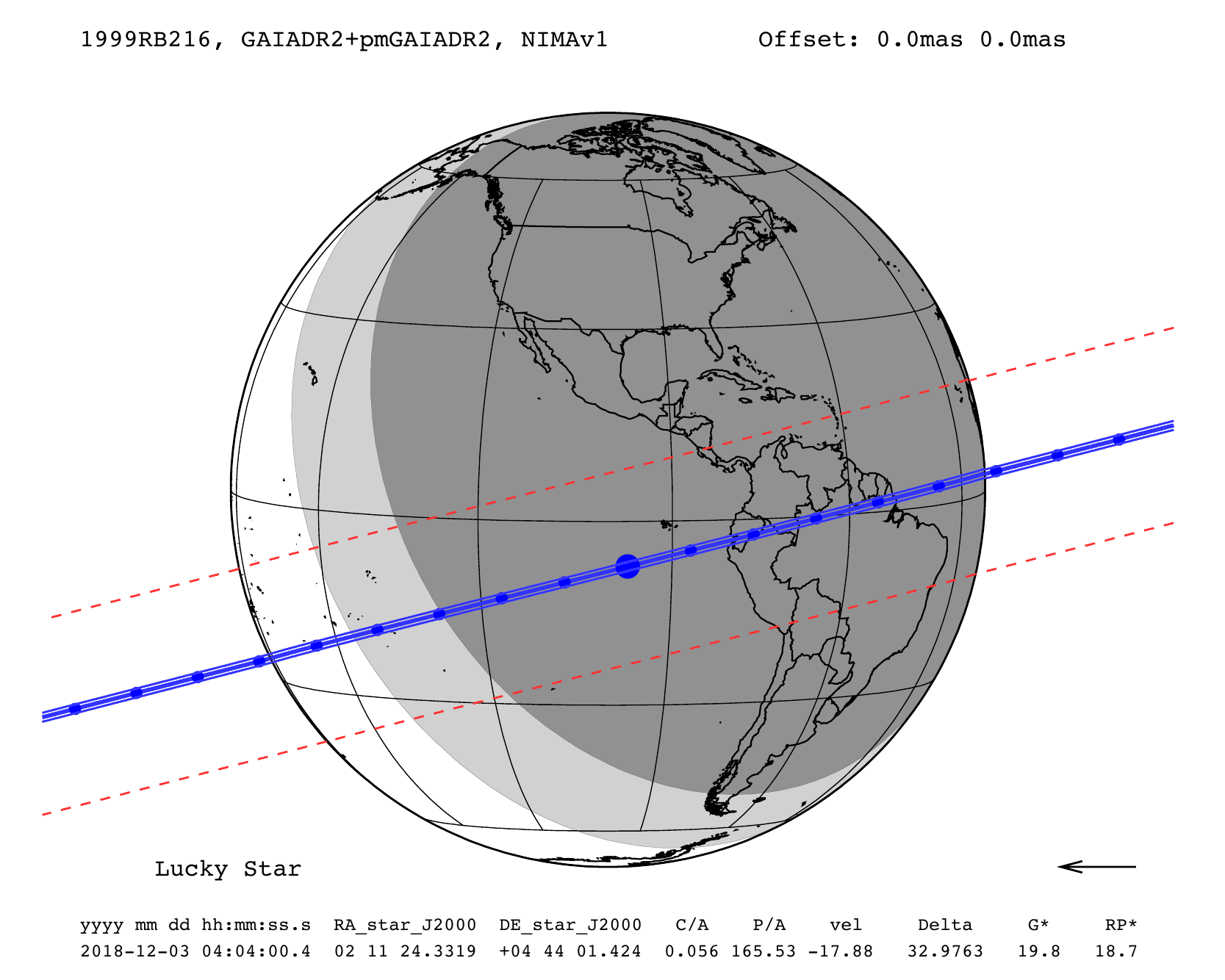}{0.6\textwidth}{(b)}}

\caption{Example of prediction result and orbit refinement for TNO (137295) 1999 RB216 Panel (a): Same as Fig.~\ref{fig:occmaps2} for the TNO (137295) 1999 RB216. The ephemeris JPL\#18 is used to determine the differences NIMA minus JPL. This object belongs to the TNO main group, Table~\ref{tab:selkbores}. Panel (b): occultation map, telling date and time (UTC) of the closest approach (largest blue point) between the shadow path and the geocenter, equatorial coordinates of the candidate star to be occulted, the closest approach (angular distance as seen from the occulting body, in arcsec, between the geocenter and the largest blue dot), the position angle (angle measured, in degrees, from the north pole to segment linking the geocenter and the largest blue point, counted clockwise), an estimate of the shadow speed on the Earth (km s$^{-1}$), the geocentric distance of the occulting body (AU), the Gaia DR2 G magnitude of the occulted star normalized to a reference shadow speed of 20 km s$^{-1}$, and the magnitude of the occulted star from the Gaia DR2 red photometer also normalized to the same reference shadow speed. Dark and white areas indicate nighttime and daylight, respectively. The gray zone shows the limits of the terminator \citep[see also][for a detailed description]{2010A&A...515A..32A}. The distance between the blue lines indicates the diameter of the occulting body. The prediction uncertainty is given by the red dashed lines. The arrow in the bottom right corner of the map indicates the sense of the shadow's movement.
\label{fig:occmaps1}}
\end{figure*}

These maps can be found from the address \url{http://lesia.obspm.fr/lucky-star/des/predictions} along with a link to specific ongoing campaigns where intense astrometric efforts are done to orbit improvement. These specific campaigns are those for which worldwide alerts are sent. The basic pieces of information given by the maps are as illustrated by Fig.~\ref{fig:occmaps1} (b).

Prediction maps, plots with ephemeris uncertainties, as well as the respective ephemerides (bsp files) are available and are constantly updated at the websites mentioned earlier in the text.

		\section{Comments and conclusions}

We used 4,292,847 individual CCD frames from the DES collaboration to search for all known small bodies in the solar system. They represent a huge amount of high quality data, obtained by a single instrument and treated in a homogeneous and reproducible way. 

Our procedure provided accurate positions from the DECam images and can be extended to other detectors. The correction for the chromatic refraction is a step to profit from the full excellence in space metrology of the instrument. Such a correction is in progress.

The whole procedure, from image retrieval from the DES database to the prediction of stellar occultations, is part of a pipeline that is being implemented in a high performance computational environment. Nevertheless, we interfered a number of times to check the data quality. As a result, the pipeline itself is refined.

The accuracy of the positions has a stronger dependence on the objects' magnitude than on its number of observations. This means that the low detection threshold adopted by PRAIA software to extract the faintest sources did not compromise the quality of the results.

Our detection efficiency is around 90\% to $r<$22 and we detect objects as faint as $r\sim$24, more than 1 magnitude fainter than the average limiting magnitude in the same band. Again, this indicates that the faintest sources were found.

The basic results provided here (astrometry, orbits, and predictions to TNOs and Centaurs) are constantly updated as more observations from the DES or from other telescopes become available, the LSST being a natural continuation of this work. These results are available in dedicated websites.

\section{Acknowledgments}

M.V.B.H. acknowledges CAPES fellowship. J.I.B.C. acknowledges CNPq grant 308150/2016-3. MA thanks to the CNPq (Grants 473002/2013-2 and 308721/2011-0) and FAPERJ (Grant E-26/111.488/2013). RV-M thanks grants: CNPq-304544/2017-5, 401903/2016-8, Faperj: PAPDRJ-45/2013 and E-26/203.026/2015. F.B-R. acknowledges CNPq grant 309578/2017-5. M.M.G. acknowledges a Capes fellowship (Proc. n.º 88887.144443/2017-00). The work leading to these results has received funding from the National Institute of Science and Technology of the e-Universe project (INCT do e-Universo, CNPq grant 465376/2014-2). The work leading to these results has received funding from the European Research Council under the European Community's H2020 2014-2020 ERC grant Agreement n$^{\rm o}$ 669416 ``Lucky Star''. 

Funding for the DES Projects has been provided by the U.S. Department of Energy, the U.S. National Science Foundation, the Ministry of Science and Education of Spain, the Science and Technology Facilities Council of the United Kingdom, the Higher Education Funding Council for England, the National Center for Supercomputing Applications at the University of Illinois at Urbana-Champaign, the Kavli Institute of Cosmological Physics at the University of Chicago, the Center for Cosmology and Astro-Particle Physics at the Ohio State University, the Mitchell Institute for Fundamental Physics and Astronomy at Texas A\&M University, Financiadora de Estudos e Projetos, Funda{\c c}{\~a}o Carlos Chagas Filho de Amparo {\`a} Pesquisa do Estado do Rio de Janeiro, Conselho Nacional de Desenvolvimento Cient{\'i}fico e Tecnol{\'o}gico and the Minist{\'e}rio da Ci{\^e}ncia, Tecnologia e Inova{\c c}{\~a}o, the Deutsche Forschungsgemeinschaft and the Collaborating Institutions in the Dark Energy Survey. 

The Collaborating Institutions are Argonne National Laboratory, the University of California at Santa Cruz, the University of Cambridge, Centro de Investigaciones Energ{\'e}ticas, Medioambientales y Tecnol{\'o}gicas-Madrid, the University of Chicago, University College London, the DES-Brazil Consortium, the University of Edinburgh, the Eidgen{\"o}ssische Technische Hochschule (ETH) Z{\"u}rich, 
Fermi National Accelerator Laboratory, the University of Illinois at Urbana-Champaign, the Institut de Ci{\`e}ncies de l'Espai (IEEC/CSIC), the Institut de F{\'i}sica d'Altes Energies, Lawrence Berkeley National Laboratory, the Ludwig-Maximilians Universit{\"a}t M{\"u}nchen and the associated Excellence Cluster Universe, the University of Michigan, the National Optical Astronomy Observatory, the University of Nottingham, The Ohio State University, the University of Pennsylvania, the University of Portsmouth, SLAC National Accelerator Laboratory, Stanford University, the University of Sussex, Texas A\&M University, and the OzDES Membership Consortium.

Based in part on observations at Cerro Tololo Inter-American Observatory, National Optical Astronomy Observatory, which is operated by the Association of 
Universities for Research in Astronomy (AURA) under a cooperative agreement with the National Science Foundation.

The DES data management system is supported by the National Science Foundation under Grant Numbers AST-1138766 and AST-1536171. The DES participants from Spanish institutions are partially supported by MINECO under grants AYA2015-71825, ESP2015-66861, FPA2015-68048, SEV-2016-0588, SEV-2016-0597, and MDM-2015-0509, some of which include ERDF funds from the European Union. IFAE is partially funded by the CERCA program of the Generalitat de Catalunya.
Research leading to these results has received funding from the European Research
Council under the European Union's Seventh Framework Program (FP7/2007-2013) including ERC grant agreements 240672, 291329, and 306478. We  acknowledge support from the Australian Research Council Centre of Excellence for All-sky Astrophysics (CAASTRO), through project number CE110001020, and the Brazilian Instituto Nacional de Ci\^encia  e Tecnologia (INCT) e-Universe (CNPq grant 465376/2014-2).

This manuscript has been authored by Fermi Research Alliance, LLC under Contract No. DE-AC02-07CH11359 with the U.S. Department of Energy, Office of Science, Office of High Energy Physics. The United States Government retains and the publisher, by accepting the article for publication, acknowledges that the United States Government retains a non-exclusive, paid-up, irrevocable, world-wide license to publish or reproduce the published form of this manuscript, or allow others to do so, for United States Government purposes.

Special thanks to J. Giorgini (JPL – Pasadena – California) for his help with Horizons ephemerides.

	\appendix

\section{Astrometric results}

\begin{longrotatetable}
\startlongtable
\begin{deluxetable*}{l|rrcccccccr}
\tablecaption{Statistics from the reduction of TNOs and Centaurs - main sources.
\label{tab:selkbores}}
\tablehead{
\colhead{Object} & \colhead{App. Mag.} & \colhead{RA-$3\sigma$} & \colhead{DE-$3\sigma$} & \colhead{$\sigma_{\alpha}{\rm cos}\delta$} & \colhead{$\sigma_{\delta}$} & \colhead{Exposure} & \colhead{Positions} & \colhead{Detections} & \colhead{Images} & \colhead{Filters} \\
\colhead{Id.} & \colhead{\tablenotemark{a}} & \multicolumn2c{(mas)\tablenotemark{b}} & \multicolumn2c{(mas)} & \colhead{min. (s) max.} & \colhead{} & \colhead{}
}
\colnumbers
\startdata
TNO & & & & & & & & & &\\
\cline{1-1} & & & & & & & & & &\\
 1999 OZ3                                         &        23.1     (0.1)     &      2015   &      928  &       69 &     68  &   200 200     &     6  &    6  &    6   &    6r                         \\
 2001 QP297                                       &        23.2     (0.3)     &      2605   &     1889  &      152 &    152  &    90 90      &     4  &    4  &   15   &    2r1i                       \\
 2001 QQ297                                       &        23.19    (0.06)    &      2761   &     2168  &      285 &    134  &    90 90      &     6  &    6  &   20   &    2r                         \\
 2001 QQ322                                       &        22.8     (0.2)     &      3154   &     1736  &      180 &    165  &    90 90      &    13  &   13  &   18   &    4r4i2z                     \\
 2001 QS322                                       &        23.1     (0.1)     &      1774   &     1399  &       91 &    103  &    90 90      &    12  &   13  &   21   &    6r2i2z                     \\
 2003 QQ91                                        &        23.4     (0.1)     &      1802   &     1662  &       83 &    174  &    90 90      &     6  &    6  &   16   &    1r1i                       \\
 2003 QT91                                        &        23.5     (0.1)     &      3074   &     2141  &       60 &     89  &    90 90      &     4  &    4  &   15   &    1r1i                       \\
 2003 QV90                                        &        22.9     (0.1)     &      4122   &     2598  &      196 &    124  &    90 90      &     3  &    3  &   17   &    1i                         \\
 2003 QY111                                       &        23.3     (0.4)     &      3386   &     2144  &      191 &    219  &    90 90      &     5  &    5  &   17   &    2r                         \\
 2003 QZ111                                       &        23.2     (0.1)     &      4725   &     2510  &      172 &     63  &    90 90      &    11  &   11  &   19   &    3r3i                       \\
 2003 SQ317                                       &        23.0     (0.1)     &      4030   &     1745  &       98 &     94  &    90 90      &    10  &   14  &   19   &    3g4r1i1z                   \\
 2003 SR317                                       &        23.2     (0.1)     &       438   &      311  &      174 &    115  &    90 90      &     4  &    4  &   15   &    1r1i                       \\
 2003 UJ292                                       &        22.6     (0.4)     &       474   &      294  &      135 &     90  &    90 90      &     5  &    5  &    9   &    2i2z                       \\
 2004 SC60                                        &        22.886   (0.008)   &       177   &      151  &       39 &     53  &    90 90      &     7  &    7  &    7   &    2g3r1i1z                   \\
 2006 QF181                                       &        23.31    (0.09)    &       258   &      196  &       73 &    118  &    90 90      &     4  &    4  &   22   &    2r1i                       \\
 2006 QQ180                                       &        23.3     (0.1)     &      1373   &      973  &      116 &     98  &    90 90      &    15  &   15  &   19   &    1g4r2i5z                   \\
 2006 UO321                                       &        23.5     (0.1)     &       333   &      279  &      274 &    204  &    90 90      &    10  &   10  &   22   &    1g2r2i                     \\
 2007 TD418                                       &        24.27    (0.06)    &      2190   &      738  &      154 &    123  &    90 200     &    26  &   29  &  133   &    4g6r4i2z                   \\
 2007 TZ417                                       &        23.7     (0.2)     &      1356   &     1598  &       56 &    276  &    90 90      &    14  &   14  &   31   &    4g5r1i                     \\
 2010 RD188                                       &        22.17    (0.02)    &      1718   &     1630  &      429 &    209  &    90 90      &    13  &   13  &   13   &    3g4r3i3z                   \\
 2010 RF188                                       &        23.4     (0.1)     &       437   &      285  &      262 &     58  &    90 90      &    10  &   10  &   12   &    1g3r4i2z                   \\
 2010 RF64                                        &        21.5     (0.1)     &      2213   &     1188  &      175 &     94  &    90 90      &    11  &   11  &   16   &    3g3r3i1z                   \\
 2010 RO64                                        &        22.12    (0.05)    &       141   &      128  &       37 &     43  &    90 90      &     4  &    4  &   10   &    2g1r1i                     \\
 2010 TJ                                          &        22.00    (0.04)    &      1854   &     1785  &      102 &     95  &    90 90      &    13  &   14  &   15   &    2g3r2i4z                   \\
 2010 TY53                                        &        20.90    (0.07)    &       138   &      176  &       37 &     13  &    90 90      &    19  &   20  &   20   &    6g7r1i5z                   \\
 2012 TC324                                       &        22.81    (0.06)    &       122   &      103  &       97 &    119  &    90 90      &    24  &   24  &   26   &    5g3r5i6z                   \\
 2012 TD324                                       &        23.1     (0.1)     &       708   &      444  &      260 &    181  &    90 90      &     9  &    9  &   14   &    4g1r1i2z                   \\
\cellcolor{blue!25}  2012 YO9                     &        23.6     (0.2)     &      1711   &     1759  &      169 &    174  &    90 200     &    22  &   25  &  174   &    5r2i                       \\
\cellcolor{blue!25}  2013 QP95                    &        23.4     (0.1)     &       144   &      261  &       93 &     67  &    90 400     &   203  &  218  &  321   &    20g21r40i84z               \\
\cellcolor{blue!25}  2013 RB98                    &        23.5     (0.1)     &       870   &     1004  &      190 &    117  &    90 200     &    51  &   53  &   92   &    4g11r12i13z                \\
\cellcolor{blue!25}  2013 RD98                    &        24.13    (0.06)    &       314   &      399  &      163 &    137  &    90 400     &   165  &  188  &  655   &    4g25r32i19z                \\
\cellcolor{blue!25}  2013 RR98                    &        23.85    (0.02)    &      3450   &     3244  &       98 &    129  &    90 90      &    14  &   16  &   30   &    2g2r4i5z                   \\
\cellcolor{blue!25}  2013 SE99                    &        24.0     (0.1)     &       982   &     1195  &      226 &    232  &   150 400     &    30  &   46  &  479   &    3i                         \\
 2013 SZ99                                        &        23.6     (0.2)     &       458   &      357  &      273 &    349  &    90 90      &     6  &    6  &   19   &    1r1i                       \\
\cellcolor{blue!25}  2013 TH159                   &        24.2     (0.2)     &      5163   &     3873  &      171 &    158  &   200 400     &    41  &   60  &  670   &    1g7r1i                     \\
 2013 TM159                                       &        23.3     (0.2)     &       727   &      486  &      129 &    122  &    90 90      &    17  &   17  &   24   &    2g3r4i3z                   \\
 2013 UK15                                        &        23.2     (0.1)     &      4669   &     2236  &      248 &     58  &    90 90      &     3  &    3  &    6   &    1r1i                       \\
 2013 UO15                                        &        22.9     (0.1)     &       320   &      254  &       56 &     96  &    90 90      &     4  &    4  &   10   &    1r1i                       \\
 2013 UQ15                                        &        23.440   (0.004)   &       473   &      387  &      120 &     86  &    90 90      &     5  &    5  &   11   &    2g3r                       \\
 2013 UR15                                        &        23.7     (0.2)     &       492   &      336  &      168 &     77  &    90 90      &     6  &    6  &   16   &    1g1r2i                     \\
 2014 GE54                                        &        22.81    (0.07)    &       151   &      128  &       41 &     43  &   150 150     &    20  &   21  &   35   &    6g6r4i3z                   \\
 2014 LO28                                        &        21.69    (0.08)    &       213   &      107  &       30 &     37  &    90 90      &    13  &   13  &   14   &    5g3r3i1z                   \\
 2014 OD394                                       &        22.93    (0.08)    &      3146   &      663  &       56 &     40  &    90 90      &     6  &    6  &   14   &    1g2r2i1z                   \\
 2014 OQ394                                       &        22.29    (0.09)    &       152   &      114  &       55 &     79  &    90 90      &     7  &    7  &    8   &    3r2i1z                     \\
 2014 OR394                                       &        22.7     (0.1)     &       241   &      165  &      100 &    185  &    90 90      &     4  &    4  &    5   &    1r1i1z                     \\
 2014 QA442                                       &        21.113   (0.003)   &       272   &      290  &      114 &     52  &    90 90      &    10  &   10  &   26   &    2g4r2i2z                   \\
 2014 QC442                                       &        23.3     (0.1)     &      2772   &      683  &      363 &     58  &    90 90      &     4  &    5  &    9   &    1g1r1i1z                   \\
\cellcolor{blue!25}  2014 QE442                   &        23.73    (0.05)    &      4058   &     2250  &       90 &    104  &    90 200     &    10  &   17  &   37   &    3g1r4i                     \\
\cellcolor{blue!25}  2014 QF442                   &        23.8     (0.3)     &      5780   &     5998  &      149 &    151  &    90 90      &    13  &   14  &   25   &    3g4r2i                     \\
\cellcolor{blue!25}  2014 QG442                   &        23.03    (0.05)    &      3160   &     3239  &      140 &    272  &    90 90      &    14  &   14  &   22   &    4g4r6i                     \\
\cellcolor{blue!25}  2014 QL441                   &        22.8     (0.2)     &      2024   &      683  &      136 &    130  &    90 200     &    73  &  102  &  111   &    11g14r16i29z               \\
\cellcolor{blue!25}  2014 QM441                   &        23.5     (0.2)     &      1225   &      470  &      152 &    110  &    90 200     &    70  &   86  &  153   &    6g15r21i21z                \\
\cellcolor{blue!25}  2014 QR441                   &        23.7     (0.1)     &      1625   &     2738  &      130 &     87  &    90 200     &    83  &   93  &  177   &    11g22r19i17z               \\
\cellcolor{blue!25}  2014 QU441                   &        26.0     (0.1)     &      5857   &     3317  &      117 &    104  &    90 200     &    27  &   29  &  106   &    1g8r5i1z                   \\
 2014 SK349                                       &        22.7     (0.1)     &       248   &      190  &       33 &     54  &    90 90      &    21  &   22  &   22   &    6g5r6i4z                   \\
\cellcolor{blue!25}  2014 SQ350                   &        24.00    (0.08)    &      3726   &     2697  &      112 &    126  &    90 400     &    44  &   53  &  208   &    5g9r12i2z                  \\
\cellcolor{blue!25}  2014 SZ348                   &        24.44    (0.08)    &       911   &     1313  &      131 &    115  &    90 400     &   197  &  239  &  515   &    7g36r52i45z                \\
\cellcolor{blue!25}  2014 TT85                    &        23.6     (0.2)     &       620   &      845  &      168 &    149  &    90 200     &    31  &   39  &  260   &    6r6i                       \\
\cellcolor{blue!25}  2014 UF224                   &        24.1     (0.1)     &      1050   &     1558  &      158 &    163  &    90 400     &    84  &  105  &  499   &    3g15r14i7z                 \\
\cellcolor{blue!25}  2014 UZ224                   &        23.75    (0.02)    &      2322   &     3439  &       89 &     75  &    90 90      &    13  &   13  &   19   &    2g4r5i1z                   \\
\cellcolor{blue!25}  2014 XY40                    &        23.01    (0.05)    &      2780   &     2735  &      138 &     87  &    90 90      &    13  &   13  &   15   &    3g4r3i3z                   \\
\cellcolor{blue!25}  2015 PD312                   &        23.6     (0.1)     &      5337   &     2765  &      172 &    125  &    90 200     &    16  &   21  &   55   &    1g4r4i                     \\
\cellcolor{blue!25}  2015 PF312                   &        22.82    (0.07)    &      1832   &      726  &       75 &     68  &    90 200     &    37  &   39  &   57   &    8g10r8i7z                  \\
\cellcolor{blue!25}  2015 RR245                   &        22.624   (0.001)   &       118   &       90  &       19 &     54  &    90 90      &     5  &    5  &    6   &    2g2r1i                     \\
 2015 RT245                                       &        22.9     (0.1)     &      1582   &     1000  &      113 &     70  &    90 90      &     9  &    9  &   16   &    3r4i2z                     \\
 2015 RU245                                       &        23.9     (0.2)     &      4749   &      799  &       88 &    111  &    90 90      &     9  &   13  &   20   &    1g2r2i                     \\
 2015 RW245                                       &        23.11    (0.09)    &      5624   &     5717  &      655 &    445  &    90 90      &     6  &    6  &   16   &    1r1i1z                     \\
 2015 TS350                                       &        23.06    (0.09)    &       866   &     2444  &       51 &     79  &    90 90      &     7  &    8  &   11   &    1g1r4z                     \\
\cellcolor{blue!25}  2015 UK84                    &        23.22    (0.08)    &      5265   &     5214  &       40 &     98  &    90 90      &    14  &   14  &   19   &    2g3r4i5z                   \\
(119956)  2002 PA149                              &        23.2     (0.1)     &      2322   &     1471  &      177 &    139  &    90 90      &    11  &   11  &   16   &    1g2r4i2z                   \\
(120348)  2004 TY364                              &        21.01    (0.09)    &       182   &      209  &       71 &     28  &    90 90      &    19  &   19  &   19   &    4g6r4i4z                   \\
(134210)  2005 PQ21                               &        23.5     (0.1)     &      1828   &     1179  &       89 &     50  &    90 90      &     8  &    8  &   18   &    1g2r2i                     \\
(136199) Eris  2003 UB313                         &        19.05    (0.02)    &       120   &      109  &        8 &      8  &    90 90      &    21  &   22  &   22   &    6g3r5i7z                   \\
(137295)  1999 RB216                              &        23.1     (0.1)     &       777   &      382  &      143 &     82  &    90 90      &    25  &   25  &   26   &    6g6r6i6z                   \\
(139775)  2001 QG298                              &        22.5     (0.2)     &       127   &      101  &       44 &     46  &    90 90      &    14  &   14  &   14   &    4g4r2i4z                   \\
(143707)  2003 UY117                              &        20.97    (0.08)    &       128   &       99  &       20 &     82  &    90 90      &     3  &    3  &    4   &    1i2z                       \\
(144897)  2004 UX10                               &        21.00    (0.02)    &        75   &       76  &       48 &     20  &    90 90      &     6  &    6  &    6   &    1g3r1i1z                   \\
(145452)  2005 RN43                               &        20.36    (0.08)    &       104   &       77  &        8 &     10  &    90 90      &    11  &   11  &   11   &    3g2r2i4z                   \\
(145474)  2005 SA278                              &        22.6     (0.1)     &       153   &      138  &       43 &     25  &    90 90      &    19  &   19  &   22   &    7g4r5i2z                   \\
(145480)  2005 TB190                              &        21.65    (0.09)    &       108   &       73  &       79 &     30  &    90 90      &    19  &   19  &   19   &    6g4r4i4z                   \\
(184212)  2004 PB112                              &        23.9     (0.2)     &      1376   &     1124  &      122 &    101  &    90 90      &     8  &    8  &   10   &    1g2r3i1z                   \\
(303775)  2005 QU182                              &        21.26    (0.03)    &       140   &      128  &       88 &     50  &    90 90      &     9  &    9  &   10   &    1g1r3i4z                   \\
(307616)  2003 QW90                               &        22.25    (0.03)    &        99   &       76  &       31 &     50  &    90 90      &    16  &   17  &   19   &    4g3r5i4z                   \\
(309239)  2007 RW10                               &        21.67    (0.07)    &        89   &       91  &       47 &     20  &    90 90      &    16  &   17  &   17   &    4g3r5i4z                   \\
(385191)  1997 RT5                                &        23.3     (0.2)     &      2466   &     1746  &      116 &     58  &    90 90      &     7  &    7  &   16   &    3r4i                       \\
(385199)  1999 OE4                                &        23.16    (0.05)    &       594   &      498  &       60 &     31  &   200 200     &     6  &    6  &    6   &    6r                         \\
(385201)  1999 RN215                              &        22.9     (0.3)     &      1936   &     1607  &      127 &    247  &    90 90      &     6  &    6  &   17   &    3r2i                       \\
(385458)  2003 SP317                              &        23.49    (0.04)    &      2827   &     1930  &      101 &     34  &    90 90      &     6  &    6  &   21   &    2r1z                       \\
\cellcolor{blue!25} (437360)  2013 TV158          &        22.8     (0.1)     &       101   &      121  &       52 &     46  &    90 400     &   438  &  467  &  504   &    44g72r101i214z             \\
(44594)  1999 OX3                                 &        20.972   (0.005)   &        90   &       66  &       14 &     20  &    90 90      &     9  &    9  &    9   &    2g3r2i2z                   \\
\cellcolor{blue!25} (451657)  2012 WD36           &        24.0     (0.1)     &       376   &      416  &      156 &    129  &    90 200     &    46  &   51  &  195   &    4g12r6i2z                  \\
(455171)  1999 OM4                                &        23.2     (0.1)     &       584   &      574  &       32 &     76  &   200 200     &     6  &    6  &    6   &    6r                         \\
(469372)  2001 QF298                              &        22.0     (0.1)     &       123   &       97  &       34 &     49  &    90 90      &    16  &   16  &   16   &    4g4r4i4z                   \\
(469750)  2005 PU21                               &        23.24    (0.07)    &       137   &       97  &      122 &     73  &    90 90      &    19  &   20  &   21   &    4g4r4i5z                   \\
(47171) Lempo  1999 TC36                          &        20.59    (0.04)    &        76   &       66  &       32 &     18  &    90 90      &    11  &   12  &   12   &    2g2r4i3z                   \\
\cellcolor{blue!25} (471954)  2013 RM98           &        22.4     (0.2)     &       121   &       92  &       79 &     67  &    90 150     &    18  &   18  &   21   &    7g3r5i3z                   \\
\cellcolor{blue!25} (472262)  2014 QN441          &        22.8     (0.2)     &       113   &      148  &       60 &     40  &    90 200     &    90  &   94  &  109   &    14g19r20i36z               \\
(480017)  2014 QB442                              &        23.3     (0.1)     &       148   &      125  &       64 &     42  &    90 90      &    19  &   21  &   26   &    3g6r5i5z                   \\
\cellcolor{blue!25} (483002)  2014 QS441          &        22.2     (0.2)     &       586   &      601  &       70 &     85  &    90 200     &    26  &   27  &   49   &    6g6r5i9z                   \\
\cellcolor{blue!25} (491767)  2012 VU113          &        24.0     (0.2)     &       285   &      404  &      181 &    108  &    90 200     &    40  &   47  &  107   &    3g14r7i4z                  \\
\cellcolor{blue!25} (491768)  2012 VV113          &        23.6     (0.1)     &       448   &      545  &      163 &    139  &    90 200     &    32  &   38  &  196   &    4r8i2z                     \\
\cellcolor{blue!25} (495189)  2012 VR113          &        23.3     (0.1)     &       359   &      310  &       97 &     92  &    90 200     &    74  &   78  &  114   &    7g16r17i21z                \\
\cellcolor{blue!25} (495190)  2012 VS113          &        23.5     (0.1)     &       515   &      426  &      119 &     68  &    90 400     &   191  &  200  &  254   &    24g32r52i70z               \\
\cellcolor{blue!25} (495297)  2013 TJ159          &        23.2     (0.1)     &      2334   &     1310  &       77 &     58  &    90 150     &    17  &   18  &   24   &    1g3r5i4z                   \\
(503883)  2001 QF331                              &        23.458   (0.007)   &       344   &      271  &      138 &     89  &    90 90      &    11  &   11  &   17   &    2g3r3i1z                   \\
(504555)  2008 SO266                              &        22.3     (0.2)     &       121   &      134  &       38 &     51  &    90 90      &    19  &   19  &   19   &    6g6r4i3z                   \\
(504847)  2010 RE188                              &        22.8     (0.1)     &       286   &      198  &       99 &     45  &    90 90      &     4  &    4  &    5   &    1g1r1i1z                   \\
\cellcolor{blue!25} (505412)  2013 QO95           &        23.4     (0.1)     &       439   &      371  &       76 &    105  &    90 200     &    46  &   52  &   81   &    7g10r12i12z                \\
(505446)  2013 SP99                               &        23.3     (0.2)     &       416   &      280  &      178 &    132  &    90 90      &     8  &    8  &   17   &    2r2i                       \\
(505447)  2013 SQ99                               &        23.2     (0.1)     &       322   &      239  &      129 &     57  &    90 90      &    10  &   10  &   19   &    1g2r4i                     \\
(505448)  2013 SA100                              &        23.4     (0.3)     &       352   &      258  &       65 &     83  &    90 90      &    15  &   15  &   20   &    2g4r6i2z                   \\
(505476)  2013 UL15                               &        23.8     (0.2)     &       392   &      253  &       78 &     69  &    90 90      &     4  &    4  &   11   &    1g3r                       \\
(508338)  2015 SO20                               &        22.5     (0.1)     &       127   &      105  &       49 &     27  &    90 90      &    17  &   19  &   20   &    4g4r5i3z                   \\
(87555)  2000 QB243                               &        23.8     (0.1)     &      2956   &     2056  &       68 &    102  &    90 90      &     6  &    6  &    7   &    1g2r2i1z                   \\
\cline{1-1} & & & & & & & & & &\\
Centaur & & & & & & & & & &\\
\cline{1-1} & & & & & & & & & &\\
2004 DA62                      &       23.30    (0.03)    &    1471   &     4914   &      90  &    36   &   90 90    &       4  &    4  &   15  &     2r2i              \\
2007 UM126                     &       22.5     (0.1)     &    4629   &      815   &      83  &    54   &   90 90    &      18  &   19  &   22  &     7g4r5i1z          \\
2011 SO277                     &       23.3     (0.1)     &     420   &      368   &      45  &    64   &   90 90    &      16  &   17  &   19  &     4g1r6i4z          \\
2012 PD26                      &       22.72    (0.09)    &    1591   &      783   &     311  &   172   &   90 90    &      11  &   13  &   15  &     5g2r1i2z          \\
\cellcolor{blue!25}2013 RG98   &       23.3     (0.1)     &     264   &      780   &      85  &    78   &   90 400   &     207  &  224  &  271  &     23g31r41i97z1Y    \\
2014 OX393                     &       22.70    (0.06)    &     779   &      452   &      86  &    62   &   90 90    &       4  &    4  &    5  &     1g1r1i1z          \\
\cellcolor{blue!25}2014 QO441  &       23.63    (0.06)    &     254   &      350   &      87  &    97   &   90 400   &     119  &  145  &  301  &     13g19r31i34z      \\
\cellcolor{blue!25}2014 QP441  &       23.8     (0.3)     &     866   &      593   &     111  &    95   &   90 400   &      74  &  123  &  436  &     2g17r17i11z       \\
\cellcolor{blue!25}2014 SB349  &       23.80    (0.04)    &    2793   &     1964   &     110  &    39   &   90 200   &      12  &   18  &   62  &     2g3r4i2z          \\
2014 SS303                     &       22.07    (0.05)    &    5559   &     2124   &      71  &    52   &   90 90    &       4  &    5  &    7  &     1r1i2z            \\
2015 RV245                     &       23.61    (0.07)    &    2878   &     3169   &     260  &    23   &   90 90    &       4  &    4  &    7  &     2g1r1i            \\
2015 VV1                       &       21.77    (0.03)    &      85   &       87   &      27  &    36   &   90 90    &       4  &    4  &    4  &     1g2r1i            \\
(2060) Chiron  1977 UB         &       18.5     (0.2)     &      53   &       50   &      18  &    13   &   90 90    &       7  &    7  &    7  &     3g1r1i2z          \\
(472265) 2014 SR303            &       22.0     (0.2)     &      95   &       79   &      27  &    30   &   90 90    &      18  &   19  &   22  &     6g6r3i3z          \\
\enddata
\tablecomments{Column (1): Object identification. Those discovered by the DES are highlighted. Column (2): Average magnitude as obtained from the bluest filter. Columns (3) and (4): 3$\sigma$ uncertainty in the ephemeris position in RA and Dec, respectively. Columns (5) and (6): standard deviations as obtained from the observed positions minus those from the respective JPL ephemeris, in RA and Dec respectively. Column (7): minimum and maximum exposure times of the images from which a position was obtained. Columns (8),(9), and (10): number of positions obtained, number of detections delivered by the astrometric code (all positions, no eliminations), and total number of images with exposure times greater than or equal to 50 seconds, respectively. Column (11): Number of magnitudes per filter found to a given object in the DES database. Note that the total number of filters in each row of column (11) is always less than or equal to the respective number of positions in Column (8). This is because either a magnitude was not found in the DES database for a given position or the position itself was not found in the DES database.}
\tablenotetext{a}{Bluest magnitude from the DES. If no magnitude from the DES is available, V magnitude given the the JPL - Horizons System - is used.}
\tablenotetext{b}{As provided by the JPL, Horizons System.}
\end{deluxetable*}
\end{longrotatetable}
\begin{longrotatetable}
\startlongtable
	\begin{deluxetable*}{l|rrcccccccr}
	\tablecaption{Statistics from the reduction of TNOs and Centaurs - extension sources.
	\label{tab:kboextend}}
	\tablehead{
\colhead{Object} & \colhead{App. Mag.} & \colhead{RA-$3\sigma$} & \colhead{DE-$3\sigma$} & \colhead{$\sigma_{\alpha}{\rm cos}\delta$} & \colhead{$\sigma_{\delta}$} & \colhead{Exposure} & \colhead{Positions} & \colhead{Detections} & \colhead{Images} & \colhead{Filters} \\
\colhead{Id.} & \colhead{\tablenotemark{a}} & 			\multicolumn2c{(mas)\tablenotemark{b}} & \multicolumn2c{(mas)} & \colhead{min. (s) max.} & \colhead{} & \colhead{}
}
	\colnumbers
	\startdata
TNO & & & & & & & & & &\\
\cline{1-1} & & & & & & & & & &\\
(160091) 2000 OL67                         &               23.2     (0.2)    &    15828   &     7732   &      42  &    61  &    90 90     &      6  &    6  &   16  &   2r2i1z      \\ 
\cellcolor{blue!25} 2013 RP98              &               23.58   (0.08)    &    20450   &     5490   &      57  &    62  &    90 90     &      7  &    7  &   15  &   2g1r3i1z    \\ 
\cellcolor{blue!25} 2013 RQ98              &               23.0     (0.2)    &    27691   &    13113   &      80  &   115  &    90 90     &      7  &   11  &   29  &   3r2i        \\ 
(160256) 2002 PD149                        &               23.6     (0.2)    &    17727   &     8159   &     150  &    90  &    90 90     &      7  &    7  &   14  &   1g2r1i1z    \\ 
2003 QX111                                 &               23.0     (0.2)    &     9090   &     3775   &      98  &   106  &    90 90     &      9  &   11  &   19  &   2r4i3z      \\ 
\cellcolor{blue!25} 2014 SR350             &               23.1     (0.1)    &    20122   &     7973   &      97  &    88  &    90 90     &      9  &   12  &   26  &   4r3i        \\ 
\cellcolor{blue!25} 2015 PL312             &               23.94   (0.08)    &    30292   &    15722   &     112  &   169  &    90 400    &      9  &   23  &  199  &   3r          \\ 
\cellcolor{blue!25} 2014 UY224             &               23.53   (0.06)    &     9915   &     9898   &     103  &   126  &    90 90     &     12  &   12  &   19  &   2g4r3i1z    \\ 
2014 UC225                                 &               23.39   (0.09)    &    11304   &     6057   &     128  &    97  &    90 90     &     13  &   13  &   21  &   3g5r3i2z    \\ 
\cellcolor{blue!25} 2014 UN225             &               23.1     (0.1)    &    32391   &    24659   &      43  &    49  &    90 90     &     14  &   16  &   17  &   4g3r4i2z    \\ 
\cellcolor{blue!25} 2014 VW37              &               23.3     (0.1)    &     3657   &     7506   &     120  &    93  &    90 90     &     18  &   18  &   21  &   4g3r5i4z    \\ 
\cellcolor{blue!25} 2013 RF98              &               24.1     (0.1)    &     6582   &     6114   &      87  &   109  &   200 400    &     30  &   55  &  301  &   5r7i1z      \\ 
\cline{1-1} & & & & & & & & & &\\
Centaur & & & & & & & & & &\\
\cline{1-1} & & & & & & & & & &\\
 2013 PQ37                 &            19.93    (0.06)   &    31300   &  12480  &  0.053  & 0.016  &  90 90     &    7  &  7 &   7   &  2r2i3z  \\
\enddata
\tablecomments{Same as for Table~\ref{tab:selkbores}.}
\end{deluxetable*}
\end{longrotatetable}


\begin{longrotatetable}
\startlongtable
	\begin{deluxetable*}{l|rrcccccccr}
	\tablecaption{Statistics from the reduction of TNOs and Centaurs - doubtful sources.
	\label{tab:kbodoubt}}
	\tablehead{
\colhead{Object} & \colhead{App. Mag.} & \colhead{RA-$3\sigma$} & \colhead{DE-$3\sigma$} & \colhead{$\sigma_{\alpha}{\rm cos}\delta$} & \colhead{$\sigma_{\delta}$} & \colhead{Exposure} & \colhead{Positions} & \colhead{Detections} & \colhead{Images} & \colhead{Filters} \\
\colhead{Id.} & \colhead{\tablenotemark{a}} & 			\multicolumn2c{(mas)\tablenotemark{b}} & \multicolumn2c{(mas)} & \colhead{min. (s) max.} & \colhead{} & \colhead{}
}
	\colnumbers
	\startdata
TNO & & & & & & & & & &\\
\cline{1-1} & & & & & & & & & &\\
1996 RR20                                   &            22.802   (0.006) &       7994   &       3676    &     221  &   177  &    90 90    &       4  &    4  &   15   &    2i1z       \\ 
1999 RG215                                  &            23.7     (0.2)   &       2919   &       1919    &          &        &    90 90    &       1  &    1  &    9   &    1r         \\ 
1999 RK215                                  &            24.23            &       2590   &       2135    &          &        &    90 90    &       1  &    1  &   15   &               \\ 
2000 PC30                                   &            23.8     (0.2)   &      47499   &      19797    &     174  &    74  &   200 200   &       4  &    4  &    6   &    3r         \\ 
2000 PY29                                   &            23.9     (0.2)   &       9129   &       4247    &      97  &    97  &   200 200   &       4  &    4  &    6   &    1r         \\ 
2000 QD226                                  &            23.65            &   $>10^{6}$  &  $>10^{6}$    &          &        &    90 90    &       1  &    1  &   21   &               \\ 
2001 QH298                                  &            22.88    (0.09)  &       1824   &       1851    &          &        &    90 90    &       1  &    1  &   16   &    1g         \\ 
2001 QO297                                  &            23.6     (0.2)   &      22524   &       9941    &     154  &   148  &    90 90    &       3  &    3  &   19   &    1g1r1i     \\ 
2002 PD155                                  &            23.53            &      20548   &      11298    &          &        &    90 90    &       1  &    1  &   22   &               \\ 
2002 PG150                                  &            21.61    (0.07)  &   $>10^{6}$  &  $>10^{6}$    &          &        &    90 90    &       1  &    1  &   13   &    1z         \\ 
2002 PK149                                  &            22.48    (0.09)  &   $>10^{6}$  &  $>10^{6}$    &          &        &    90 90    &       1  &    1  &   12   &    1g         \\ 
2003 QB91                                   &            23.1     (0.1)   &      11356   &       5215    &     493  &    51  &    90 90    &       4  &    4  &   22   &    1r1i       \\ 
2005 PE23                                   &            26.93            &   $>10^{6}$  &  $>10^{6}$    &          &        &    90 90    &       1  &    1  &   21   &               \\ 
2005 PP21                                   &            22.88            &   $>10^{6}$  &  $>10^{6}$    &          &        &    90 90    &       1  &    1  &   11   &               \\ 
2005 SE278                                  &            22.19    (0.07)  &       1897   &       1498    &      47  &    30  &    90 90    &       2  &    2  &    3   &    1i1z       \\ 
2006 QC181                                  &            22.00    (0.05)  &   $>10^{6}$  &   $>10^{6}$   &          &        &    90 90    &       1  &    1  &   15   &    1g         \\ 
2006 QD181                                  &            22.88            &   $>10^{6}$  &  $>10^{6}$    &          &        &    90 90    &       1  &    1  &   13   &               \\ 
2006 QZ180                                  &            23.59            &   $>10^{6}$  &  $>10^{6}$    &          &        &    90 90    &       1  &    1  &   23   &               \\ 
2008 UA332                                  &            23.03    (0.08)  &   $>10^{6}$  &   $>10^{6}$   &          &        &    90 90    &       1  &    1  &   17   &    1g         \\ 
2010 JH124                                  &            23.2     (0.1)   &      20165   &       1810    &     798  & 1.050  &    90 150   &       3  &    3  &   43   &    1r1i       \\ 
2013 KZ18                                   &            21.65            &        136   &        104    &          &        &    90 90    &       1  &    1  &    4   &    1z         \\ 
\cellcolor{blue!25} 2013 RO98               &            22.74    (0.08)  &    $>10^{6}$ &    $>10^{6}$  &      44  &    72  &    90 90    &      16  &   16  &   18   &    4g4r4i4z   \\  
2013 UP15                                   &            24.06            &        370   &        260    &          &        &    90 90    &       1  &    1  &    7   &               \\ 
\cellcolor{blue!25} 2013 VD24               &            24.6     (0.2)   &     107390   &      54984    &     113  &   172  &   330 400   &       5  &    9  &  408   &    1r         \\ 
2013 VJ24                                   &            23.90            &   $>10^{6}$  &   $>10^{6}$   &          &        &    90 90    &       1  &    1  &   15   &               \\ 
2014 NB66                                   &            22.86    (0.02)  &        217   &        115    &      50  &    69  &    90 90    &       2  &    2  &    4   &    2g         \\ 
2014 PR70                                   &            22.98    (0.07)  &        226   &        136    &     175  &   190  &    90 90    &       2  &    2  &    2   &    1g1z       \\ 
\cellcolor{blue!25} 2014 RS63               &            22.62    (0.06)  &      85434   &      52414    &      61  &   131  &    90 90    &       6  &    6  &   13   &    3i1z       \\ 
\cellcolor{blue!25} 2014 SN350              &            22.87    (0.09)  &      28123   &      37105    &     169  &   147  &    90 90    &       6  &    6  &   21   &    3r1i       \\ 
\cellcolor{blue!25} 2014 SO350              &            24.0     (0.2)   &      91741   &      33931    &     229  &    67  &    90 90    &       7  &    8  &   23   &    1g2r3i     \\ 
2014 TB86                                   &            23.2     (0.1)   &     165206   &      42723    &      68  &    69  &    90 90    &       9  &   11  &   23   &    1g3r2i2z   \\  
\cellcolor{blue!25} 2014 TE86               &            23.2     (0.3)   &      23334   &      38078    &     190  &   140  &    90 90    &       9  &    9  &   19   &    2g3r1i     \\ 
\cellcolor{blue!25} 2014 TF86               &            23.5     (0.2)   &      47829   &      27854    &     150  &   119  &    90 90    &      12  &   12  &   26   &    1g5r2i2z   \\  
\cellcolor{blue!25} 2014 TU85               &            23.38    (0.02)  &     860527   &     132551    &     256  &    95  &    90 200   &       4  &    4  &   46   &    2r1i       \\ 
\cellcolor{blue!25} 2014 UA225              &            23.37    (0.06)  &     441418   &     196089    &      66  &    89  &    90 90    &      11  &   11  &   22   &    3g2r3i3z   \\  
\cellcolor{blue!25} 2014 UB225              &            22.74    (0.05)  &   $>10^{6}$  &    $>10^{6}$  &      57  &    41  &    90 90    &       7  &    7  &   10   &    3r1i2z     \\ 
\cellcolor{blue!25} 2014 VT37               &            24.06    (0.09)  &     196302   &      93079    &     106  &   123  &   150 200   &      11  &   11  &   74   &    2g3r1i     \\ 
\cellcolor{blue!25} 2014 YL50               &            23.4     (0.1)   &      43878   &      78972    &     129  &   178  &    90 90    &      12  &   12  &   14   &    4g3r3i1z   \\  
2014 XZ40                                   &            23.53    (0.02)  &      72326   &      56308    &      52  &   128  &    90 90    &       5  &    5  &   18   &    2g2r1i     \\ 
\cellcolor{blue!25} 2015 PK312              &            25.01            &    $>10^{6}$ &     224438    &     330  & 1.328  &    90 330   &       3  &    3  &  133   &               \\ 
\cellcolor{blue!25} 2015 QT11               &            24.3     (0.2)   &     465277   &     218826    &     188  &   179  &   150 400   &       9  &   11  &  239   &    1g2i       \\ 
2015 RS245                                  &            24.05            &      46613   &       5685    &     833  &   929  &    90 200   &       4  &    4  &   87   &               \\ 
2015 RX245                                  &            24.35            &       1320   &       1461    &     161  &   446  &    90 90    &       2  &    2  &   28   &               \\ 
2015 SV20                                   &            22.56            &    $>10^{6}$ &    $>10^{6}$  &          &        &    90 90    &       1  &    1  &   13   &               \\ 
2015 TN178                                  &            21.4     (0.5)   &        175   &        641    &     267  &   371  &    90 90    &       2  &    2  &    2   &    2i         \\ 
\cellcolor{blue!25} 2016 QP85               &            23.6     (0.2)   &    $>10^{6}$ &     306619    &     660  &   236  &    90 90    &       3  &    3  &   13   &    1r         \\ 
(148112) 1999 RA216                         &            22.7     (0.1)   &       2402   &       1746    &     190  &   216  &    90 90    &       2  &    2  &   12   &    1i         \\ 
(307982) 2004 PG115                         &            20.63    (0.01)  &        132   &         77    &          &        &    90 90    &       1  &    1  &    1   &    1r         \\ 
(474640) 2004 VN112                         &            23.42            &        748   &        816    &          &        &    90 90    &       1  &    1  &    1   &               \\ 
(501581) 2014 OB394                         &            21.42    (0.03)  &        136   &         97    &      37  &    37  &    90 90    &       2  &    2  &    2   &    1i1z       \\ 
(506121) 2016 BP81                          &            23.2     (0.1)   &        397   &        276    &      74  &   137  &    90 90    &       2  &    2  &    3   &    1g1i       \\ 
\cline{1-1} & & & & & & & & & &\\
Centaur & & & & & & & & & &\\
\cline{1-1} & & & & & & & & & &\\
2007 VL305                    &        22.7     (0.1)    &     11377  &      2924   &     266  &   245   &   90 90     &      3   &   3   &   7   &    1r             \\
2011 OF45                     &        21.12    (0.04)   &       565  &       334   &          &         &   90 90     &      1   &   1   &   1   &    1z             \\
2013 RN30                     &        22.6     (0.2)    &   7971879  &   5872040   &     516  & 1.049   &   90 90     &      3   &   3   &  22   &    2g1z           \\
2013 SV99                     &        24.1     (0.1)    &   2145099  &   1273035   &     192  &   151   &   90 400    &     16   &  20   &  55   &    4g2r5i         \\
2013 TS20                     &        21.83    (0.03)   &  36400409  &  14095067   &          &         &   90 90     &      1   &   1   &   6   &    1g             \\
2014 SW223                    &        21.83    (0.05)   &       762  &       545   &          &         &   90 90     &      1   &   1   &   1   &    1i             \\
2014 TK34                     &        21.14    (0.03)   &       310  &       197   &          &         &   90 90     &      1   &   1   &   1   &    1i             \\
\cellcolor{blue!25}2015 VT152 &        23.5     (0.1)    &   1501899  &   1234480   &     112  &    41   &   90 90     &      6   &   9   &  16   &    1g2r3i         \\
2016 VF1                      &        21.71    (0.04)   &  53403030  & 272552177   &     644  & 1.161   &   90 90     &      3   &   3   &  20   &    1i             \\
(471339) 2011 ON45            &        21.04    (0.07)   &       115  &        74   &          &         &   90 90     &      1   &   1   &   1   &    1z             \\
\enddata
\tablecomments{Same as for Table~\ref{tab:selkbores}.}
\end{deluxetable*}
\end{longrotatetable}

\bibliography{references}
\end{document}